\documentclass[reprint,pre,twocolumn,footinbib,aps,superscriptaddress,nobalancelastpage]{revtex4-2}

\usepackage[usenames, dvipsnames ]{xcolor}
\usepackage{amsmath, amsfonts, amssymb, amsthm, amstext, bm, bbm, multirow}
\usepackage{physics}
\usepackage[colorlinks=true, linkcolor=blue, citecolor=red, 
urlcolor=blue, anchorcolor = blue, filecolor = blue, hypertexnames=true]{hyperref}
\usepackage{graphicx, caption, subcaption}
\usepackage{ragged2e}
\DeclareCaptionJustification{justified}{\justifying}
\usepackage{float}

\newcommand{\Fig}[1]{Fig.~\ref{#1}}
\newcommand{\Figs}[2]{Figs.~\ref{#1} and \ref{#2}}
\newcommand{\Eq}[1]{Eq.~(\ref{#1})}
\newcommand{\Eqs}[2]{Eqs.~(\ref{#1}) and~(\ref{#2})}

\newcommand{\apjl}{Astrophys. J. Lett.}
\newcommand{\apjs}{Astrophys. J. Suppl. Ser.}

\begin{document}


\title{Quantum Effects on Dynamic Structure Factors in Dense Magnetized Plasmas}

\author{Tushar Mondal}
 \email{tushar.mondal@icts.res.in}
\affiliation{
 International Centre for Theoretical Sciences, Tata Institute of Fundamental Research, Bengaluru 560089, India
}
\author{Gianluca Gregori}
\email{gianluca.gregori@physics.ox.ac.uk}
\affiliation{
 Department of Physics, University of Oxford, Parks Road, Oxford OX1 3PU, United Kingdom
}


\begin{abstract}

We extend the classical magnetohydrodynamics formalism to include nonlocal quantum behavior via the phenomenological Bohm potential. We then solve the quantum magnetohydrodynamics equations to obtain a new analytical form of the dynamic structure factor (DSF), a fundamental quantity linking theory and experiments. Our results show that the three-peak structure---one central Rayleigh peak and two Brillouin peaks---of the DSF arising from quantum hydrodynamic fluctuations becomes (in general) a five-peak structure---one central Rayleigh peak and two pairs of peaks associated with fast and slow magnetosonic waves. The Bohm contribution influences the positions and characteristics (height, width, and intensity) of the peaks by introducing three significant modifications: (a) an increase in effective thermal pressure, (b) a reduction in the adiabatic index, and (c) an enhancement of effective thermal diffusivity. The multiple DSF peaks enable concurrent measurements of diverse plasma properties, transport coefficients, and thermodynamic parameters in magnetized dense plasmas. The potential for experimental validation of our theory looms large, particularly through future experiments conducted at state-of-the-art laser facilities.

\end{abstract}

\maketitle

\section{\label{sec:intro}Introduction}

In the Universe, matter is often found in extreme states, with densities comparable to solids and temperatures of a few electron volts, known as warm dense matter (WDM). As the temperature rises to the order of a few keV, one enters the regime of a hot dense plasma. Matter under such extreme conditions is characterized by partially degenerate electrons and strongly correlated ions \cite{1982RvMP...54.1017I, 2011RvMP...83..885S}. Precise mesurements of plasma conditions, including transport and thermodynamic properties in both WDM and dense plasmas, are of high importance for understanding high energy density physics phenomena. The applications span a wide range, from modeling the atmosphere of neutron stars \cite{2009ApJ...703..994D} and magnetars \cite{2006RPPh...69.2631H, 2021ApJ...913L..12M} to investigating the interiors of giant planets \cite{1999Sci...286...72G, 2018NatAs...2..452S}, white \cite{2001ApJ...549L.219B, 2013PhRvL.110g1102D} and brown dwarfs \cite{2014ApJS..215...21B}, as well as for the advancement of inertial confinement fusion \cite{2004PhPl...11..339L, 2018HEDP...28....7G}, with its promise of potentially abundant and clean energy for the future. However, the extreme conditions pose significant challenges in diagnostics, often preventing direct measurement of even basic plasma properties.

A fundamental quantity that describes the microscopic space- and time-dependent behavior of WDM and dense plasmas is the dynamic structure factor (DSF) \cite{hansen2013theory}. The DSF contains essential information on the energy and angular distribution of the scattering that results from the individual and collective behavior of electrons and ions. This property, in turn, makes the DSF a function of macroscopic plasma quantities such as density, temperature and magnetic fields. Since the DSF is directly proportional to the X-ray Thomson scattering cross-section, it can be probed through laser plasma experiments \cite{2009RvMP...81.1625G}. The comparison between experimental and calculated DSF not only validates theoretical models but also allows the simultaneous measuremet of plasma properties that are otherwise challenging to determine. Thus, the DSF serves as a powerful diagnostic tool for probing, understanding, and characterizing the intricate behavior of WDM and dense plasmas, making a significant contribution to the advancement of high-energy density physics research.

The DSF is formally defined as the Fourier transform in space and time of the density autocorrelation function. For density fluctuations, the hydrodynamic description is highly successful due to its analytical solvability \cite{hansen2013theory}, connecting fundamental thermodynamics and transport properties of plasmas in a simple physical form. These analytical results have shown good agreement with molecular dynamic simulations \cite{2011PhRvE..83a5401M, 2011PhRvE..84d6401M} in predicting the dynamical response of strongly coupled classical plasmas. However, classical hydrodynamics falls short in capturing various astrophysical and laboratory conditions, such as systems with dynamically dominant magnetic fields and quantum effects. In the presence of a background magnetic field, previous results have been limited to special cases like weakly coupled plasmas \cite{1961PhRv..122.1663S, sheffield2010plasma} or those neglecting particle correlations \cite{1979PhRvD..19.2868H}. Recently, Bott and Gregori \cite{2019PhRvE..99f3204B} computed the DSF in a magnetized, high-density plasma, describing collective excitations through magnetohydrodynamics (MHD).

In this paper, we extend the classical MHD formalism to incorporate quantum effects through the introduction of the Bohm potential \cite{1952PhRv...85..166B, haas2011quantum, 2019SciA....5.1634L, 2019PhPl...26i0601B}. This novel approach allows us to derive an analytical expression for the modified DSF in strongly coupled, partially degenerate, and magnetized plasmas. Our comprehensive framework encompasses finite viscosity, thermal conductivity, electrical resistivity, and quantum nonlocality. We demonstrate that quantum effects do not change the number of excitation modes compared to their classical counterparts. However, we will show that the introduction of quantum contributions leads to substantial modifications in the positions and characteristics of these DSF resonances through (a) a significant enhancement of effective thermal pressure, (b) a reduction in the adiabatic index, and (c) an augmentation of effective thermal diffusivity.

Traditionally, the Landau-Placzek ratio \cite{hansen2013theory}, defined as the ratio of the Rayleigh peak intensity to that of the two Brillouin peaks, offers a means to estimate the specific heat ratio (adiabatic index) of fluids across a wide range of thermodynamic conditions \cite{1966JChPh..44.2785C, 1967JChPh..47...31O, 2011JChPh.135m4510P, 2017PhRvE..96d2608Z}. In quantum MHD, we demonstrate that the expression for the Landau-Placzek ratio remains unchanged as $R_{\text{LP}} = \gamma' - 1$, with the adiabatic index $\gamma$ being modified due to the Bohm contributions. Additionally, we introduce another significant parameter, the `F-to-S ratio,' which quantifies the intensity ratio of the fast magnetosonic wave peak to the slow magnetosonic wave peak. This ratio provides an experimental avenue for measuring the magnetic field strength within the plasma medium.

The paper is organized as follows. In Section~\ref{sec:mhd}, we present the governing equations of MHD in a standard form, incorporating quantum dynamics through the Bohm potential. Section~\ref{sec:dsf} provides a derivation of the density autocorrelation function, and consequently, the dynamic structure factor, within the context of quantum MHD for small-amplitude fluctuations. Moving on to Section~\ref{sec: parallel fluctuations}, we derive the general form of the dynamic structure factor for fluctuations with wave vectors parallel to the magnetic field. This scenario closely resembles quantum hydrodynamics fluctuations. In Section~\ref{sec: oblique fluctuations}, we extend our analysis to derive the general form of the dynamic structure factor for oblique fluctuations within the quantum MHD framework. Finally, the paper concludes with a summary in Section~\ref{sec: conclusions}.

\section{\label{sec:mhd}General magnetohydrodynamics equations}

We first write down the general set of single-fluid MHD equations for an electron-ion plasma in the presence of heat conduction and quantum effects. The governing equations for the conservation of mass, momentum, magnetic flux, and internal energy are given, respectively, by \cite{2014ApJ...795...59C}
\begin{subequations}	
	\begin{eqnarray}
	\frac{d\rho}{dt} & = & -\rho\div \bm{u}, \label{eq:continuity} \\
	\rho \frac{d\bm{u}}{dt} & = & -\grad{p}-\div \bm{\Pi}+\frac{(\curl{\bm{B}})\times\bm{B}}{\mu_{0}}+\bm{\Phi}_{\text{Bohm}}, \label{eq:momentum} \\
	\frac{d\bm{B}}{dt} & = & (\bm{B}\cdot\bm{\nabla})\bm{u}-\bm{B}(\div\bm{u})-\curl{(\eta\curl{\bm{B}})}, \label{eq:induction} \\
	\rho\frac{d\epsilon}{dt} & = & -p\div{\bm{u}}-\bm{\Pi} :\bm{\nabla}\bm{u}+\eta\frac{\vert\curl{\bm{B}}\vert^{2}}{\mu_{0}}-\div{\bm{q}} \nonumber \\
	&& \, +\bm{\Phi}_{\text{Bohm}}\cdot\bm{u}, \label{eq:energy}
	\end{eqnarray}
\end{subequations}
where $\rho$ is the mass density, $t$ the time, $\bm{u}$ the bulk fluid velocity, $p$ the pressure, $\bm{\Pi}$ the viscosity tensor, $\bm{\Phi}_{\text{Bohm}}$ the quantum Bohm potential, $\bm{B}$ the magnetic field, $\mu_0$ the permeability of free space, $\eta$ the magnetic diffusivity, $\epsilon$ the internal energy, and $\bm{q}$ is the heat flux. Here, the convective derivative can be written as $d/dt \equiv \partial/\partial t + \bm{u} \cdot \nabla$. The expressions for the viscosity tensor, heat flux and the quantum Bohm potential are explicitly given by \cite{2014ApJ...795...59C}
\begin{subequations}
	\label{constiteqns}
	\begin{eqnarray}
	& \boldsymbol{\Pi}  =  -\zeta_s \left[\nabla \boldsymbol{u} + \left(\nabla  \boldsymbol{u}\right)^{\rm T} - \frac{2}{3} \left(\nabla \cdot \boldsymbol{u} \right) \mathbf{I} \right] 
	- \zeta_b \left(\nabla \cdot \boldsymbol{u} \right) \mathbf{I} , \qquad \; \\
	& \boldsymbol{q}  =  - \kappa \nabla T , \qquad \; \\
	& \bm{\Phi}_{\text{Bohm}}  =  \frac{\hbar^{2}\rho}{2m_{e}m_{i}}\grad{\left(\frac{\laplacian{\sqrt{\rho}}}{\sqrt{\rho}}\right)},
	\end{eqnarray}
\end{subequations}
where $\zeta_s$ is the shear viscosity (or the first coefficient of viscosity), $\zeta_b$ the bulk viscosity (or the second coefficient of viscosity), $\mathbf{I}$ the identity tensor, $\kappa$ the thermal conductivity, $T$ the fluid temperature, $\hbar$ the reduced Planck's constant, $m_e$ the electron mass, and $m_i$ the ion mass. Note that the momentum associated with a fluid element can change not only by the pressure gradient and the inertial term but also because of viscous drag, the Lorentz force, and exchange effects arising from the contribution of the Bohm potential. The energy equation correctly describes the non-local quantum effects as well as the effects arising from the nonideal heat flux.

It is worth to rewrite the internal energy conservation law (\ref{eq:energy}) as a temperature evolution equation in terms of density, bulk flow velocity, and the magnetic field. Following the first law of thermodynamics, and using thermodynamic identities and equation (\ref{eq:continuity}), we can rewrite equation~(\ref{eq:energy}) as (see Appendix~\ref{sec: Appendix-A} for details)
\begin{eqnarray}
\rho C_V {\mathrm{d} T \over \mathrm{d} t} & = & - \frac{\gamma - 1}{\alpha_T} \rho C_V \nabla \cdot \boldsymbol{u} - \boldsymbol{\Pi}:\nabla \bm{u} \nonumber\\ 
&& \, + \eta \frac{\left|\nabla \times \bm{B}\right|^2}{\mu_0} - \nabla \cdot \bm{q} +\bm{\Phi}_{\text{Bohm}}\cdot\bm{u}, \label{MHDgoveqns_temp} 
\end{eqnarray}
where $C_V$ is the heat capacity at constant volume, $\gamma$ the adiabatic index, and $\alpha_T$ the coefficient of thermal expansion. Also, we eliminate the pressure gradient in equation (\ref{eq:momentum}) using the thermodynamic identity (see Appendix~\ref{sec: Appendix-A} for details)
\begin{equation}
\nabla p = \frac{c_s^2}{\gamma} \left(\nabla \rho + \rho \alpha_T \nabla T\right), \label{thermoiden_B}
\end{equation}
where $c_s$ is the adiabatic sound speed. 

On substituting (\ref{constiteqns}) and (\ref{thermoiden_B}), we rewrite the set of MHD equations as
\begin{subequations}	
	\begin{eqnarray}
	\frac{d\rho}{dt} & = & -\rho\div \bm{u}, \label{eq:continuity_2} \\
	\rho \frac{d\bm{u}}{dt} & = & -\frac{c_s^2}{\gamma} \left(\nabla \rho + \rho \alpha_T \nabla T\right)+\frac{(\curl{\bm{B}})\times\bm{B}}{\mu_{0}} \nonumber \\
	&& \, +\bm{\Phi}_{\text{Bohm}} + \nabla(\zeta_b \div \bm{u}) \nonumber \\
	&& \, +\div \left[\zeta_s \left\{ \nabla \boldsymbol{u} + \left(\nabla  \boldsymbol{u}\right)^{\rm T} - \frac{2}{3} \left(\nabla \cdot \boldsymbol{u} \right) \mathbf{I} \right\} \right], \label{eq:momentum_2} \\
	\frac{d\bm{B}}{dt} & = & (\bm{B}\cdot\bm{\nabla})\bm{u}-\bm{B}(\div\bm{u})-\curl{(\eta\curl{\bm{B}})}, \label{eq:induction_2} \\
	\rho {\mathrm{d} T \over \mathrm{d} t} & = & - \frac{\gamma - 1}{\alpha_T} \rho \div \bm{u} - \frac{1}{C_V}\boldsymbol{\Pi}:\nabla \bm{u} \nonumber\\ 
	&& \, + \eta \frac{\left|\nabla \times \bm{B}\right|^2}{\mu_0 C_V} + \frac{1}{C_V} \div (\kappa \nabla T) \nonumber\\ 
	&& \, + \frac{1}{C_V}\bm{\Phi}_{\text{Bohm}}\cdot\bm{u}. \label{eq:energy_2}
	\end{eqnarray}
\end{subequations}
In the temperature evolution equation, we have not written down the full expression for viscous dissipation term for brevity.

\section{\label{sec:dsf}Fluctuations and the dynamic structure factor}

We now focus on the calculation of the density autocorrelation function -- and thereby the quantum-MHD dynamic structure factor in the limit of small-amplitude fluctuations. To perform this calculation, we consider small fluctuations of dynamic quantities for a fluid system in some equilibrium state, and linearize the above quantum-MHD equations. We assume the system is in equilibrium at a density $\rho_0$, bulk flow velocity $\bm{u}_0 = 0$, magnetic field $\bm{B}_0$, temperature $T_0$, sound speed $c_{s0}$, adiabatic index $\gamma_0$, coefficient of thermal expansion $\alpha_{T0}$, specific heat capacity at constant volume $C_{V0}$, thermal conductivity $\kappa_0$, bulk viscosity $\zeta_{b0}$, shear viscosity $\zeta_{s0}$, and magnetic diffusivity $\eta_0$. We then take the small-amplitude fluctuations of dynamic quantities on this equilibrium state, as 
\begin{equation}
\rho = \rho_0 + \delta \rho, \quad  \boldsymbol{u} = \delta \boldsymbol{u}, 
\quad \boldsymbol{B} = \boldsymbol{B}_0 + \delta \boldsymbol{B}, \quad T = 
T_0 + \delta T .  \label{fluctuations_def}
\end{equation}
Substituting linearization (\ref{fluctuations_def}) into equations (\ref{eq:continuity_2})--(\ref{eq:energy_2}), and neglecting terms quadratic or higher order in fluctuating quantities, we obtain
\begin{subequations}
	\label{MHDlineqns}
	\begin{eqnarray}
	\frac{\partial \delta \rho}{\partial  t}  & = & - \rho_0 \nabla \cdot \delta \boldsymbol{u}, 
	\label{MHDlineqns_dens} \\
	\rho_0 {\partial \delta \bm{u} \over \partial  t} & = & -  \frac{c_{s0}^2}{\gamma_0} \left(\nabla \delta \rho + \rho_0 \alpha_{T0} \nabla \delta T\right) \nonumber \\
	&& \, + \frac{\bm{B}_0 \cdot \nabla \delta \bm{B}}{\mu_0} -\nabla \left( \frac{\boldsymbol{B}_0 \cdot \delta \bm{B}}{\mu_0} \right) \nonumber \\
	&& \, +  \zeta_{s0} \nabla^2 \delta \bm{u} +\zeta_{c0} \nabla \left(\nabla \cdot \delta \bm{u} \right) \nonumber \\
	&& \, + \frac{\hbar^2}{4m_e m_i} \nabla \left(\nabla^2  \delta \rho \right) , \qquad \label{MHDlineqns_flow} \\ 
	{\partial \delta \boldsymbol{B} \over \partial  t}  & = & \boldsymbol{B}_0 \cdot \nabla \delta \boldsymbol{u} - \boldsymbol{B}_0 \nabla \cdot \delta \boldsymbol{u} + \eta_0 \nabla^2 \delta \boldsymbol{B},  \label{MHDlineqns_flux} \\
	{\partial \delta T \over \partial  t} & = & - \frac{\gamma_0 - 1}{\alpha_{T0}} \nabla \cdot \delta \boldsymbol{u} + \gamma_0 \chi_0 \nabla^2  \delta T ,    \label{MHDlineqns_temp}
	\end{eqnarray} 
\end{subequations}
where we have defined the `compressive' viscosity coefficient $\zeta_{c0} \equiv \zeta_{b0} - 2 \zeta_{s0}/3$, and thermal diffusivity $\chi_0 \equiv \kappa_0/\rho_0 C_{V0} \gamma_0$. Here, the terms with subscript `$0$' refer to the quantities at equilibrium.

To obtain the MHD DSF, we apply to equations (\ref{MHDlineqns_dens})--(\ref{MHDlineqns_temp}) a Laplace transform in time, and a Fourier transform in space. For any plasma quantity $\delta x$, this operation is defined as
\begin{equation*}
\widetilde{\delta x}_{k}(s) = \int_0^{\infty} \!\! \mathrm{d}t \, e^{-st}  \int_{- \infty}^{+ \infty} \! \! \mathrm{d}^3 r\, e^{i \boldsymbol{k} \cdot \boldsymbol{r}}\,\delta x(\boldsymbol{r},t),
\end{equation*}
where the quantity with tilde indicates the transformed one, and $s=\epsilon + i\omega$ is the complex Laplace variable.
Using standard properties of Fourier and Laplace transforms under derivatives, we find 
\begin{subequations}
	\begin{eqnarray}
	s\widetilde{\delta\rho}_{\bm{k}}(s) & = & -i\rho_{0}\bm{k} \cdot \widetilde{\delta\bm{u}}_{\bm{k}}(s)+\delta\rho_{\bm{k}}(0), 
	\label{transform_continuity} \\
	\rho_{0}s\widetilde{\delta\bm{u}}_{\bm{k}}(s) & = & -\frac{c_{s0}^{2}}{\gamma_{0}}\left[ i\bm{k}\widetilde{\delta\rho}_{\bm{k}}(s)+i\rho_{0}\alpha_{T0}\bm{k}\widetilde{\delta T}_{\bm{k}}(s) \right] \nonumber \\
	&& \, +i\widetilde{\delta\bm{B}}_{\bm{k}}(s)\frac{\bm{B}_{0}\cdot\bm{k}}{\mu_{0}}-i\bm{k}\frac{\bm{B}_{0}\cdot\widetilde{\delta\bm{B}}_{\bm{k}}(s)}{\mu_{0}} \nonumber \\
	&& \, -\zeta_{s0}k^{2}\widetilde{\delta\bm{u}}_{\bm{k}}(s)-\zeta_{c0}\bm{k}\left[\bm{k}\cdot\widetilde{\delta\bm{u}}_{\bm{k}}(s)\right] \nonumber \\   && \, -\frac{\hbar^{2}k^{2}}{4m_{e}m_{i}}i\bm{k}\widetilde{\delta\rho}_{\bm{k}}(s)
	+\rho_{0}\delta\bm{u_{k}}(0), \label{transform_momentum} \\	
	s \widetilde{\delta\bm{B}}_{\bm{k}}(s) & = & i(\bm{B}_{0}\cdot\bm{k})\widetilde{\delta\bm{u}}_{\bm{k}}(s) -i\bm{B}_{0} \left[\bm{k}\cdot\widetilde{\delta\bm{u}}_{\bm{k}}(s)\right] \nonumber \\
	&& \, -\eta_{0}k^{2}\widetilde{\delta\bm{B}}_{\bm{k}}(s)+\delta\bm{B}_{\bm{k}}(0), \label{transform_induction} \\	
	s\widetilde{\delta T}_{\bm{k}}(s) & = & -i\frac{\gamma_{0}-1}{\alpha_{T0}}\bm{k}\cdot\widetilde{\delta\bm{u}}_{\bm{k}}(s)-\gamma_{0}\chi_{0}k^{2}\widetilde{\delta T}_{\bm{k}}(s) \nonumber \\
	&& \, +\delta T_{\bm{k}}(0). \label{transform_temp}
	\end{eqnarray}
\end{subequations}

\subsection{\label{sec:density autocorrelation} Density autocorrelation function}

We assume that the initial density fluctuations ${\delta\rho}_{\bm{k}}(0)$ are uncorrelated with the initial velocity fluctuations ${\delta \bm{u}}_{\bm{k}}(0)$, the initial magnetic field fluctuations ${\delta \bm{B}}_{\bm{k}}(0)$, and the initial temperature fluctuations ${\delta T}_{\bm{k}}(0)$. This assumption allows ${\delta \bm{u}}_{\bm{k}}(0)$, ${\delta \bm{B}}_{\bm{k}}(0)$, and ${\delta T}_{\bm{k}}(0)$ to be set to zero. Next, we write the magnetic field fluctuations $\widetilde{\delta \boldsymbol{B}}_{\bm{k}}(s)$ and the temperature fluctuations $\widetilde{\delta T}_{\bm{k}}(s)$ in terms of velocity field fluctuations $\widetilde{\delta \bm{u}}_{\bm{k}}(s)$, using equations (\ref{transform_induction}) and (\ref{transform_temp}), respectively:
\begin{subequations}
	\begin{eqnarray}
	\widetilde{\delta\bm{B}}_{\bm{k}}(s) & = & \frac{i(\bm{k}\cdot\bm{B}_{0})\widetilde{\delta\bm{u}}_{\bm{k}}(s) - i \left[\bm{k}\cdot\widetilde{\delta\bm{u}}_{\bm{k}}(s)\right]\bm{B}_{0}}{s+\eta_{0}k^{2}}, \label{eq: magnetic field fluctuation} \\
	\widetilde{\delta T}_{\bm{k}}(s) & = & -\frac{i(\gamma_{0}-1)\left[\bm{k}\cdot\widetilde{\delta\bm{u}}_{\bm{k}}(s)\right]}{\alpha_{T0}(s+\gamma_{0}\chi_{0}k^{2})}. \label{eq: temperature fluctuation}
	\end{eqnarray}
\end{subequations}

We then solve for $\bm{B}_{0}\cdot \widetilde{\delta\bm{u}}_{\bm{k}}(s)$ in terms of $\widetilde{\delta\rho}_{\bm{k}}(s)$ and $i\rho_{0}\bm{k} \cdot \widetilde{\delta\bm{u}}_{\bm{k}}(s)$, by taking the scalar product of (\ref{transform_momentum}) with $\bm{B}_{0}$, and substituting (\ref{eq: temperature fluctuation}). This gives
\begin{multline}
\bm{B}_{0}\cdot \widetilde{\delta\bm{u}}_{\bm{k}}(s)  =  -\frac{i(\bm{k} \cdot \bm{B}_{0})}{(\rho_{0}s+\zeta_{s0}k^{2})} \Bigg[ \left( \frac{c_{s0}^{2}}{\gamma_{0}} + \frac{\hbar^{2}k^{2}}{4m_{e}m_{i}} \right) \widetilde{\delta\rho}_{\bm{k}}(s) \\
- \left\lbrace \frac{(\gamma_{0}-1)c_{s0}^{2}}{\gamma_{0}(s+\gamma_{0}\chi_{0}k^{2})} + \nu_{c0} \right\rbrace i\rho_{0}\bm{k} \cdot \widetilde{\delta\bm{u}}_{\bm{k}}(s) \Bigg]. \label{MHDlintranseqns_solveforuB_0}
\end{multline}
We subsequently evaluate $i \rho_0 \bm{k}\cdot \widetilde{\delta \bm{u}}_{\bm{k}}(s)$ in terms of $\widetilde{\delta\rho}_{\bm{k}}(s)$ alone, using the scalar product between (\ref{transform_momentum}) and $i \bm{k}$, as well as substituting (\ref{eq: magnetic field fluctuation}), (\ref{eq: temperature fluctuation}), and (\ref{MHDlintranseqns_solveforuB_0}):
\begin{widetext}
\begin{eqnarray}
i s \rho_0 \boldsymbol{k} \cdot \widetilde{\delta \boldsymbol{u}}_{\boldsymbol{k}}(s) & = & \left( \frac{k^{2}c_{s0}^{2}}{\gamma_{0}} + \frac{\hbar^{2}k^{4}}{4m_{e}m_{i}} \right) \left[1 +\frac{ \left(\boldsymbol{k} \cdot \boldsymbol{B}_0 \right)^2}{\mu_0 \left(s + \eta_0 k^2 \right) \left(\rho_ 0 s + \zeta_{s0} k^2 \right) }\right] \widetilde{\delta\rho}_{\boldsymbol{k}}(s) - 
\Bigg[\frac{\left(\gamma_0 - 1 \right) k^2 c_{s0}^2}{\gamma_{0}\left(s + \gamma_0 \chi_0 k^2 \right)} + \nu_{l0} k^2 \nonumber \\
&& + \frac{k^2 B_0^2}{\mu_0 \rho_0 \left(s + \eta_0 k^2\right)} 
+ \frac{k^2  \left(\boldsymbol{k} \cdot \boldsymbol{B}_0 \right)^2}{\mu_0 \left(s+ \eta_0 k^2\right) \left(\rho_ 0 s + \zeta_{s0} k^2\right)} \left\lbrace \frac{\left(\gamma_0 - 1 \right) c_{s0}^2}{\gamma_{0}\left(s + \gamma_0 \chi_0 k^2 \right)} + \nu_{c0} \right\rbrace \Bigg] i \rho_0 \boldsymbol{k} \cdot \widetilde{\delta \boldsymbol{u}}_{\boldsymbol{k}}(s) .
\end{eqnarray}
\end{widetext}
After some rearrangement, this gives
\begin{equation}
i\rho_{0}\bm{k} \cdot \widetilde{\delta\bm{u}}_{\bm{k}}(s) = \left[ \frac{D(k,s)}{N(k,s)}-s \right] \widetilde{\delta\rho}_{\bm{k}}(s) , \label{MHDlintranseqns_solveforku}
\end{equation}
where the functions $N(k,s)$ and $D(k,s)$ are defined as
\begin{subequations}
	\begin{eqnarray}
	N(k,s) & = & \left(s+ \gamma_0 \chi_0 k^2\right) \left(s+\nu_{l0} k^2\right) 
	\left(s+ \eta_0 k^2\right) \left(s+ \nu_{s0} k^2\right) \, \; \nonumber \\
	&& \, + \frac{\gamma_0-1}{\gamma_0} k^2 c_{s0}^2  \left(s+ \eta_0 k^2\right) \left(s+ \nu_{s0} k^2\right) 
	\nonumber \\
	&& \, + k^2 v_A^2 \left(s+ \gamma_0 \chi_0 k^2\right) \left[ s+k^2 \left( \nu_{s0}  + \nu_{c0} \cos^2{\theta} \right) \right] 
	\nonumber \\
	&& \, + \frac{\gamma_0-1}{\gamma_0} k^4 v_A^2 c_{s0}^2 \cos^2{\theta} , \\
	D(k,s) & = & \left( \frac{k^{2}c_{s0}^{2}}{\gamma_{0}} + \frac{\hbar^{2}k^{4}}{4m_{e}m_{i}} \right) \left(s+ \gamma_0 \chi_0 k^2\right) \nonumber \\
	&& \, \times \left[ \left(s+ \eta_0 k^2\right) \left(s+ \nu_{s0} k^2\right) + k^2 v_A^2 \cos^2{\theta} \right] \nonumber \\
	&& \, + s N(k,s) .\quad \label{specialfuncs}
	\end{eqnarray}
\end{subequations}
Here, we define various additional quantities: shear kinematic viscosity $\nu_{s0} \equiv \zeta_{s0}/\rho_0$, compressive kinematic viscosity $\nu_{c0} = \zeta_{c0}/\rho_0$, longitudinal kinematic viscosity $\nu_{l0} = \nu_{s0} + \nu_{c0}$, $\theta$ the angle between $\bm{B}_0$ and $\bm{k}$, and $v_A \equiv B_0/\sqrt{\mu_0 \rho_0}$ the Alfv\`en speed, where $B_0 = |\boldsymbol{B}_0|$.

Finally, we substitute (\ref{MHDlintranseqns_solveforku}) into (\ref{transform_continuity}), and solve for $\widetilde{\delta\rho}_{\boldsymbol{k}}(s)$ in terms of ${\delta\rho}_{\boldsymbol{k}}(0)$,
\begin{equation}
\widetilde{\delta\rho}_{\bm{k}}(s) = \frac{N(k,s)}{D(k,s)} \delta\rho_{\bm{k}}(0) .
\end{equation}
It provides the density autocorrelation function:
\begin{equation}
\frac{\langle\delta \rho_{\boldsymbol{k}}^*(0)\widetilde{\delta\rho}_{\boldsymbol{k}}(s)\rangle}{ \langle \delta \rho^*_{\boldsymbol{k}}(0)\delta\rho_{\boldsymbol{k}}(0)\rangle} 
= \frac{N(k,s)}{D(k,s)} . \label{autocorrdens}
\end{equation}

\subsection{\label{sec: dynamic structure factor} Dynamic structure factor}

The dynamic structure factor $S_{nn}(\boldsymbol{k},\omega)$ is \cite{2019PhRvE..99f3204B}
\begin{equation}
\label{DSFlimit}
{2\pi S_{nn}(\boldsymbol{k},\omega) \over S_{nn}(\boldsymbol{k})} = 2 \Re \left[ \lim_{\varepsilon \to 0}{\langle\delta \rho_{\boldsymbol{k}}^*(0)\widetilde{\delta\rho}_{\boldsymbol{k}}(s=\varepsilon +i \omega)\rangle \over \langle \delta \rho^*_{\boldsymbol{k}}(0)\delta\rho_{\boldsymbol{k}}(0)\rangle} \right],
\end{equation}
where 
\begin{equation}
S_{nn}(\boldsymbol{k})=\int S_{nn}(\boldsymbol{k},\omega) d\omega,
\label{eq:sum_rule}
\end{equation}
is the static structure factor.

\section{Parallel fluctuations} \label{sec: parallel fluctuations}

We first consider fluctuations whose wave vector is parallel to the magnetic field, i.e., $\cos{\theta} = 1$.  In this case, we have
\begin{subequations}
	\begin{eqnarray}
	N(k,s) & = & \left[\left(s+ \eta_0 k^2\right) \left(s+ \nu_{s0} k^2\right) + k^2 v_A^2\right] N_{\|}(k,s) , \qquad \; \\
	D(k,s) & = & \left[\left(s+ \eta_0 k^2\right) \left(s+ \nu_{s0} k^2\right) + k^2 v_A^2\right] D_{\|}(k,s) , \qquad \;
	\end{eqnarray}
\end{subequations}
where 
\begin{subequations}
	\begin{eqnarray}
	N_{\|}(k,s) & = & \left(s+ \gamma_0 \chi_0 k^2\right) \left(s+\nu_{l0} k^2\right) + \frac{\gamma_0-1}{\gamma_0} k^2 c_{s0}^2 , \qquad \\
	D_{\|}(k,s) & = & \left( \frac{k^{2}c_{s0}^{2}}{\gamma_{0}} + \frac{\hbar^{2}k^{4}}{4m_{e}m_{i}} \right) \left(s+ \gamma_0 \chi_0 k^2\right) \nonumber \\
	&& \, + s N_{\|}(k,s) . \;
	\end{eqnarray}
\end{subequations}
The density autocorrelation function becomes
\begin{equation}
\frac{\langle\delta \rho_{\boldsymbol{k}}^*(0)\widetilde{\delta\rho}_{\boldsymbol{k}}(s)\rangle}{ \langle \delta \rho^*_{\boldsymbol{k}}(0)\delta\rho_{\boldsymbol{k}}(0)\rangle} 
= \frac{N_{\|}(k,s)}{D_{\|}(k,s)} . \label{autocorrdens_par}
\end{equation}

First, neglecting all the diffusive effects, we find
\begin{subequations}
	\begin{eqnarray}
	N_{\|}(k,s) & = & s^2 + \frac{\gamma_0-1}{\gamma_0} k^2 c_{s0}^2 , \qquad \\
	D_{\|}(k,s) & = & s\left( s^2 + k^{2}c_{s0}^2 + \frac{\hbar^{2}k^{4}}{4m_{e}m_{i}} \right)   . \;
	\end{eqnarray}
\end{subequations}
The roots are then 
\begin{equation}
s_{*} = 0, \quad \pm i k c_{\text{eff}} ,
\end{equation}
where 
\begin{equation}
c_{\text{eff}} = \sqrt{c_{s0}^2 + c_{Q}^2}, \ \text{with} \ c_{Q}^2 = \frac{\hbar^{2}k^{2}}{4m_{e}m_{i}} .
\end{equation}
We then determine the density autocorrelation function in the neighbourhood of each of these roots in turn. The numerator is $N(k,s_{*}) \neq 0$ in each case. By reintroducing the diffusive terms, we have \cite{2019PhRvE..99f3204B}
\begin{itemize}
	\item \underline{$s_{*} = 0$}: let $s = \delta s \sim \chi_0 k^2, \eta_0 k^2, \nu_{s0} k^2, \nu_{l0} k^2$. Then, 
   \begin{equation*}
    D_{\|}(k,s) \approx k^2 c_{s0}^2 \, (\delta s + \chi_0 k^2) + k^2 c_{Q}^2 \, (\delta s + \gamma_0 \chi_0 k^2),
   \end{equation*}
   
	\begin{equation}
	\frac{\langle\delta \rho_{\boldsymbol{k}}^*(0)\widetilde{\delta\rho}_{\boldsymbol{k}}(s)\rangle}{ \langle \delta \rho^*_{\boldsymbol{k}}(0)\delta\rho_{\boldsymbol{k}}(0)\rangle} 
	\approx \frac{(\gamma_0-1) c_{s0}^2/(\gamma_0 c_{\text{eff}}^2)}{s+ \Gamma_{\chi} k^2} , \label{autocorrdens_neighbour_entropymode}
	\end{equation}
where
\begin{equation}
\Gamma_{\chi} = \frac{c_{s0}^2 + \gamma_0 c_Q^2}{c_{\text{eff}}^2} \chi_0 = Q\chi_0, \label{quasiperp_diffuse_SW}
\end{equation}
and
\begin{equation}
Q = \frac{c_{s0}^2 + \gamma_0 c_Q^2}{c_{\text{eff}}^2} \geq 1.
\label{eq:Q-factor}
\end{equation}
In the classical limit (i.e., $c_Q=0$), the factor $Q$ becomes unity.

	\item \underline{$s_{*} = \pm i k c_{\text{eff}} $}: let $s = \pm i k c_{\text{eff}} + \delta s$, $\delta s \sim \chi_0 k^2, \eta_0 k^2, \nu_{s0} k^2, \nu_{l0} k^2$. Then,
	\begin{equation*}
	D_{\|}(k,s) \approx -2 k^2 c_{\text{eff}}^2 \, (\delta s + \Gamma_{\|} k^2) ,
	\end{equation*}
	where
	\begin{equation}
	\Gamma_{\|} = \frac{1}{2} \left( \nu_{l0} + \gamma_0 \chi_0 - \Gamma_{\chi} \right) . 
	\label{eq:Gamma_parallel}
	\end{equation}
	It follows that
	\begin{equation}
	\frac{\langle\delta \rho_{\boldsymbol{k}}^*(0)\widetilde{\delta\rho}_{\boldsymbol{k}}(s)\rangle}{ \langle \delta \rho^*_{\boldsymbol{k}}(0)\delta\rho_{\boldsymbol{k}}(0)\rangle} 
	\approx \frac{(\gamma_0 c_Q^2 + c_{s0}^2)/(2\gamma_0 c_{\text{eff}}^2)}{s \mp i k c_{\text{eff}} + \Gamma_{\|} k^2} . 
	\end{equation}
	
\end{itemize}
The dynamic structure factor can be derived using (\ref{DSFlimit}):
\begin{widetext}
	\begin{equation}
	\label{hydrofrac}
	{2\pi S_{nn}(k, \omega) \over S_{nn}(k)} \approx {\gamma_0-1 \over \gamma_0}{c_{s0}^2 \over c_{\text{eff}}^2} \left[{2\Gamma_{\chi} k^2 \over \omega^2 +  \left(\Gamma_{\chi} k^2 \right)^2}\right]
	+ \frac{Q}{\gamma_0}\left[{\Gamma_{\|} k^2 \over \left(\Gamma_{\|} k^2\right)^2 + \left(\omega +c_{\text{eff}}k\right)^2}
	\right. + \left. {\Gamma_{\|} k^2 \over \left(\Gamma_{\|} k^2\right)^2 + \left(\omega -c_{\text{eff}}k\right)^2}\right].
	\end{equation}
\end{widetext}
\begin{figure*}
	\centering
	\begin{subfigure}[b]{0.485\textwidth}
		\centering
		\includegraphics[width=\textwidth,height=0.8\textwidth]{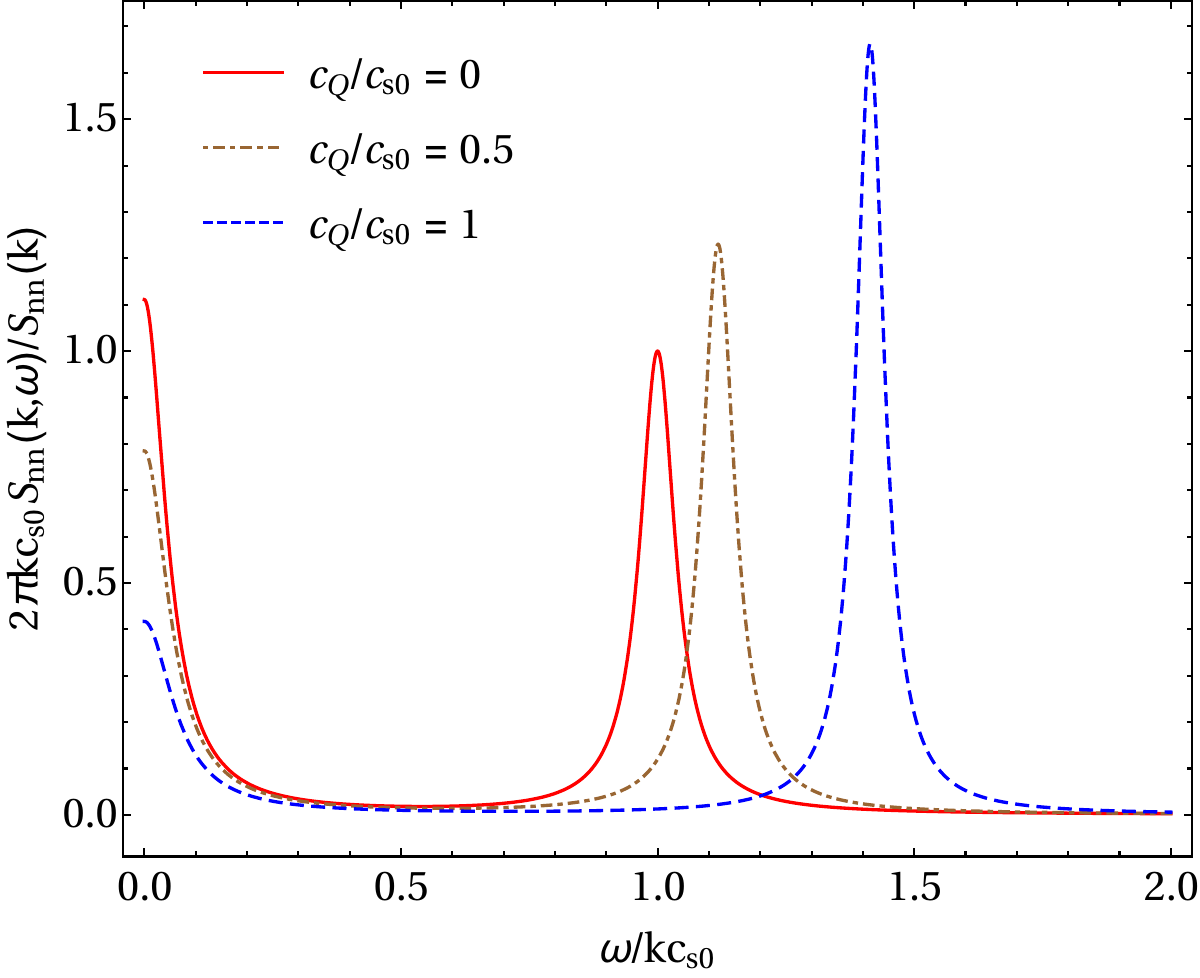}
		\caption{The dimensionless values of the dissipative terms: $k \chi_0/c_{s0} = 0.05$, $k \eta_0/c_{s0} = 0.05$, $k \nu_{s0}/c_{s0} = 0.0335$ and $k \nu_{c0}/c_{s0} = 0.0165$.}
		\label{fig:parallel_1}
	\end{subfigure}
	\hfill
	\begin{subfigure}[b]{0.485\textwidth}
		\centering
		\includegraphics[width=\textwidth,height=0.8\textwidth]{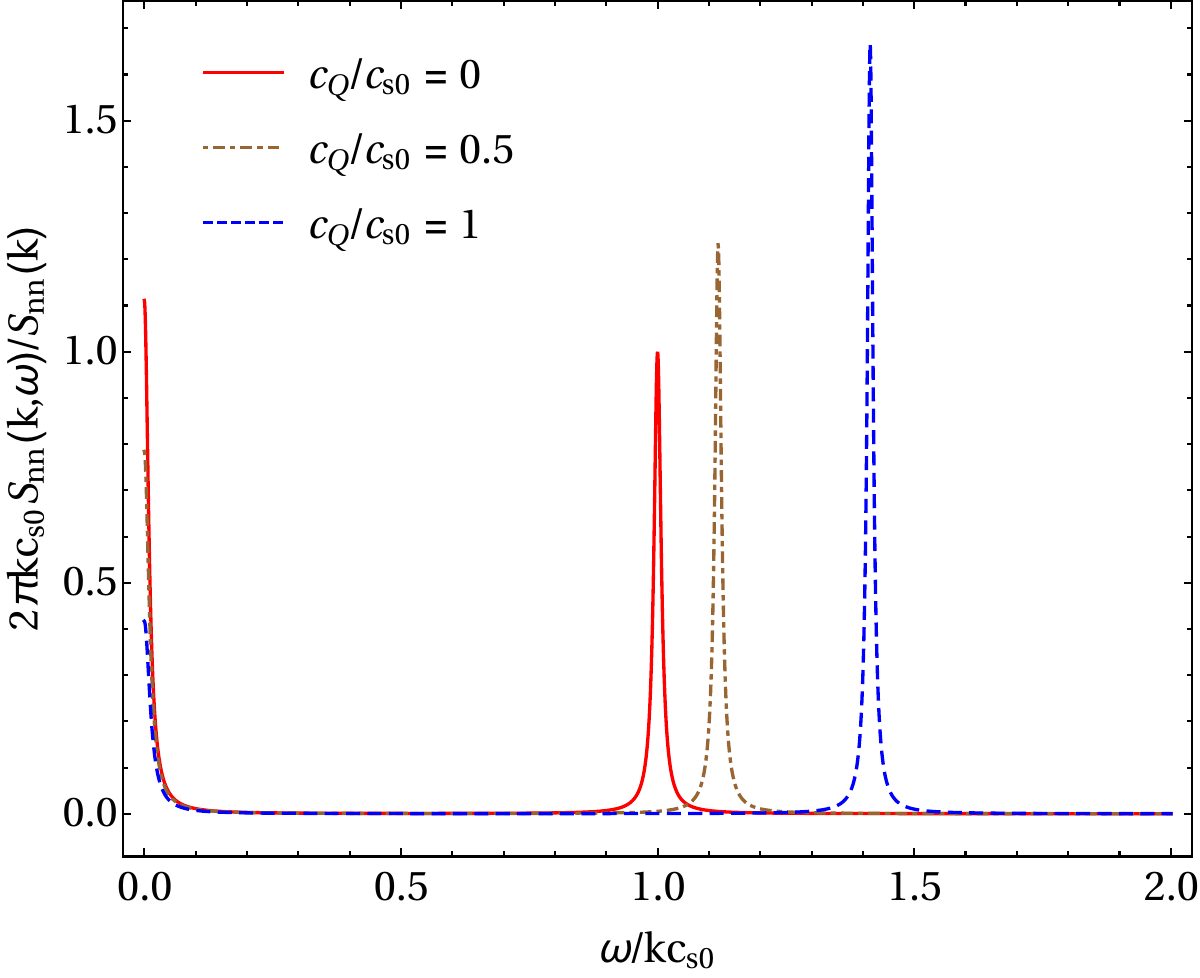}
		\caption{The dimensionless values of the dissipative terms: $k \chi_0/c_{s0} = 0.01$, $k \eta_0/c_{s0} = 0.01$, $k \nu_{s0}/c_{s0} = 0.0067$ and $k \nu_{c0}/c_{s0} = 0.0033$.}
		\label{fig:parallel_2}
	\end{subfigure}
	\caption{\textit{The dynamic structure factor in magnetized, high density plasma with increasing quantum effects at $\theta=0^{\circ}$}. The structure factor is presented in a dimensionless form; this is obtained via $s \mapsto s k c_{s0}$.  With this mapping, the magnitude of the various dissipative terms are represented by the dimensionless numbers $k \chi_0/c_{s0}$, $k \eta_0/c_{s0}$, $k \nu_{s0}/c_{s0}$ and $k \nu_{c0}/c_{s0}$. The peak magnitude is normalised to the classical $(c_Q/c_{s0}=0)$, parallel case. We choose $\gamma_0=5/3$.}
	\label{fig:parallel}
\end{figure*}
The dynamic structure factors for parallel modes are shown in Fig.~\ref{fig:parallel} for different values of $c_Q / c_{s0}$. The calculated structure factor is similar to the hydrodynamic structure factor with some modifications arising from the contribution of the Bohm potential. In MHD, parallel fluctuations of magnetic fields propagate as the Alfv$\grave{\text{e}}$n waves and do not have a density perturbation associated with them. Since the parallel compressive fluctuations do not interact with the magnetic fields, the MHD structure factor for parallel wave numbers becomes identical to the hydrodynamic one. The spectrum consits three peaks: the central Rayleigh peak at $\omega =0$, and two Brillouin peaks at $\omega = \pm k c_{\text{eff}}$. The physical nature of two different peaks can be explained qualitatively through the thermodynamic theory of fluctuations. Since the sound propagation is an adiabatic process, the density fluctuations can be decomposed into two types: entropy fluctuation at constant pressure and pressure fluctuation at constant entropy. The Rayleigh peak is associated with the nonpropagating nature of the entropy fluctuation, whereas two shifted Brillouin peaks are associated with adiabatic pressure fluctuations which propagate as sound waves. Hence, the position of Brillouin peaks is given by the dispersion relation of sound waves. The peak shapes are Lorentzian, and these are broadened due to different dissipative processes which damp out the fluctuations. 
The expressions of the full width at half maximum (FWHM) for Rayleigh and Brillouin peaks are given by, respectively, 
\begin{equation}
	W_R = 2 \frac{\Gamma_{\chi} k}{c_{s0}},\ \text{and} \ W_B = 2 \frac{\Gamma_{\|} k}{c_{s0}}.
	\label{eq:width_parallel}
\end{equation}
The corresponding peak heights are
\begin{equation}
	H_R = \frac{4}{W_R} \left(1- {Q \over \gamma_0}\right), \ \text{and} \ H_B = \frac{2}{W_B} \frac{Q}{\gamma_0}.
	\label{eq:height_parallel}
\end{equation}
The total integrated intensity of the Rayleigh peak and each of the two Brillouin peaks are
\begin{subequations}
	\begin{eqnarray}
	I_R & = & \left[{\gamma_0-1 \over \gamma_0}{c_{s0}^2 \over c_{\text{eff}}^2}\right] S_{nn}(k) = \left[ 1-\frac{Q}{\gamma_0} \right] S_{nn}(k) , \qquad \\
	I_B & = & \frac{Q}{2\gamma_0} S_{nn}(k) . \;
	\end{eqnarray}
\end{subequations}
Thus, $I_R + 2I_B = S_{nn}(k)$, which is a particular case of the sum rule (\ref{eq:sum_rule}). The ratio of the intensity of the central Rayleigh peak to that of the two shifted Brillouin peaks is given by
\begin{equation}
\frac{I_R}{2I_B} = \left(\frac{\gamma_0}{Q} -1 \right) = (\gamma'_0 -1).
\end{equation}
This is known as the Landau--Placzek ratio.

In comparison with the classical hydrodynamic case, the Bohm contribution brings a number of similarities and differences. While the frequency position of the central Rayleigh peak remains unchanged, the frequency positions of the Brillouin peaks are shifted to $\omega = k c_{\text{eff}} \gtrsim kc_{s0}$. Such modified waves are physically similar to sound waves, except for the effective equilibrium pressure being increased by additional quantum pressure arising from the nonlocal Bohm contribution. Most significantly, there are two important modifications: (a) \textit{the quantum effects enhance the effective thermal diffusivity by a factor of $Q$}, i.e., $\chi'_0 = Q\chi_0$, and (b) \textit{it reduces the adiabatic index by the same factor of $Q$}, i.e., $\gamma'_0 = \gamma_0 / Q$.

For the Rayleigh peak, thermal diffusivity alone determines the width via $\Gamma_{\chi}$. With increasing quantum effects, the FWHM of the Rayleigh peak increases due to the enhancement of the effective thermal diffusivity. The height of the Rayleigh peak decreases with increasing quantum effects because of the combined contributions of the enhanced thermal diffusivity and a reduction in the adiabatic index as per Eq.~(\ref{eq:height_parallel}). On the other hand, the width of the Brillouin peak depends on both viscosity and thermal diffusivity via $\Gamma_{\|}$, which decreases with increasing quantum effects. It further enhances the height of the Brillouin peak following Eq.~(\ref{eq:height_parallel}). These claims are illustrated in Fig.~\ref{fig:parallel}.

\section{Oblique fluctuations} \label{sec: oblique fluctuations}

Next, we focus on oblique fluctuations, where the wave vector makes an arbitary angle, $\theta$, with the magnetic field. The objective is to carry out an analytical calculation for the DSF using the quantum MHD formalism. Through this approach, we aim to gain a comprehensive understanding of the scattering spectrum, considering various combinations of oblique scattering angles, magnetic field strengths, and quantum effects. We then discuss how these factors, along with thermodynamic and transport coefficients, collectively influence the positions and shapes (height and width) of different peaks in the scattering spectra.

The positions of various peaks can be obtained from the roots of $D(k,s)$. By neglecting all the diffusive effects, we find 
\begin{equation}
D(k,s) \approx s  \left[s^4 + \left(k^2 c_{\text{eff}}^2 + k^2 v_A^2\right)s^2 + k^4 v_A^2 c_{\text{eff}}^2 \cos^2{\theta}\right] 
\, .
\end{equation}
The five roots are then
\begin{equation}
s_{*} = 0, \quad \pm i k c_F, \quad \pm i k c_S,
\end{equation}
with associated peak frequencies
\begin{equation}
\omega^2 = 0, \quad k^2 c_F^2, \quad k^2 c_S^2,
\end{equation}
where
\begin{subequations}
\begin{align}
	c_F & = \left[ \frac{1}{2} \left \lbrace \left(c_{\text{eff}}^2 + v_A^2\right) + \sqrt{\left(c_{\text{eff}}^2 + v_A^2\right)^2 - 4 c_{\text{eff}}^2 v_A^2 \cos^2{\theta}} \right \rbrace \right]^{1/2} , \label{eq:cF}  \\
	c_S & = \left[ \frac{1}{2} \left \lbrace \left(c_{\text{eff}}^2 + v_A^2\right) - \sqrt{\left(c_{\text{eff}}^2 + v_A^2\right)^2 - 4 c_{\text{eff}}^2 v_A^2 \cos^2{\theta}} \right \rbrace \right]^{1/2} . \label{eq:cS} 
\end{align}
\label{eq:cF_cS}
\end{subequations}
Using the same approach as for parallel fluctuations, an analytical form of the DSF can be derived using Eq.~(\ref{DSFlimit}):
\begin{widetext}
	\begin{eqnarray}
	\label{oblique}
	{2\pi S_{nn}(k, \omega) \over S_{nn}(k)} & \approx & \left(1-{Q \over \gamma_0}\right) \left[{2\Gamma_{\chi} k^2 \over \omega^2 +  \left(\Gamma_{\chi} k^2 \right)^2}\right]
	+ \frac{Q}{\gamma_0} \Bigg[ \left( \frac{c_F^2-v_A^2}{2c_{F}^2-v_A^2-c_{\text{eff}}^2} \right) \Bigg \lbrace {\Gamma_{F} k^2 \over \left(\Gamma_{F} k^2\right)^2 + \left(\omega +c_{F} k\right)^2} + \nonumber \\
	&& \hspace{-25pt}{\Gamma_{F} k^2 \over \left(\Gamma_{F} k^2\right)^2 + \left(\omega -c_{F} k\right)^2}  \Bigg \rbrace
	 + \left( \frac{c_S^2-v_A^2}{2c_S^2-v_A^2-c_{\text{eff}}^2} \right) \left \lbrace {\Gamma_{S} k^2 \over \left(\Gamma_{S} k^2\right)^2 + \left(\omega +c_{S}k\right)^2}
	\right. + \left. {\Gamma_{S} k^2 \over \left(\Gamma_{S} k^2\right)^2 + \left(\omega -c_{S}k\right)^2}\right \rbrace \Bigg],
	\end{eqnarray}
where
\begin{subequations}
	\begin{eqnarray}
	\Gamma_{F} & = & \frac{1}{2} \left[ \left( \frac{c_F^2-v_A^2}{2c_{F}^2-v_A^2-c_{\text{eff}}^2} \right) \left(\gamma_0-Q\right) \chi_0 + \left( \frac{c_F^2-c_{\text{eff}}^2}{2c_{F}^2-v_A^2-c_{\text{eff}}^2} \right) \eta_0 + \nu_{s0} +\frac{c_{F}^2}{c_{\text{eff}}^2} \left( \frac{c_F^2-v_A^2}{2c_{F}^2-v_A^2-c_{\text{eff}}^2} \right) \nu_{c0} \right], \label{eq:oblique_diffuse_1} \\
	\Gamma_{S} & = & \frac{1}{2} \left[ \left( \frac{c_S^2-v_A^2}{2c_{S}^2-v_A^2-c_{\text{eff}}^2} \right) \left(\gamma_0-Q\right) \chi_0 + \left( \frac{c_S^2-c_{\text{eff}}^2}{2c_{S}^2-v_A^2-c_{\text{eff}}^2} \right) \eta_0 + \nu_{s0} +\frac{c_{S}^2}{c_{\text{eff}}^2} \left( \frac{c_S^2-v_A^2}{2c_{S}^2-v_A^2-c_{\text{eff}}^2} \right) \nu_{c0} \right]. \label{eq:oblique_diffuse_2} 
	\end{eqnarray}
	\label{eq:oblique_diffuse}
\end{subequations}
\end{widetext}

For oblique fluctutations, the dynamic structure factor consists of five peaks instead of three. The central Rayleigh peak at $\omega = 0$ remains unchanged, as in the case of pure hydrodynamic or parallel fluctuations. Among the four peaks, there exist a pair of peaks at frequencies $\omega = kc_F \gtrsim kc_{\text{eff}} \gtrsim kc_{s0}$, and a new pair of peaks has emerged at frequencies $\omega = kc_S $. The FWHMs of these different peaks are given by, respectively,
\begin{eqnarray}
W_R = 2 \frac{\Gamma_{\chi} k}{c_{s0}}, 
W_F = 2 \frac{\Gamma_F k}{c_{s0}}, \ \text{and} \
W_S = 2 \frac{\Gamma_S k}{c_{s0}}.
\label{eq:oblique_width}
\end{eqnarray}
The width of both pairs of peaks depends on a linear combination of thermal diffusivity, resistivity, and the viscosities via their corresponding $\Gamma$'s. 
The heights of the corresponding peaks are given by
\begin{subequations}
	\begin{eqnarray}
	H_R &=& \frac{4}{W_R} \left(1-{Q \over \gamma_0}\right) , \label{eq:oblique_height_R} \\
	H_F &=& \frac{2}{W_F} \frac{Q}{\gamma_0} \left( \frac{c_F^2-v_A^2}{2c_{F}^2-v_A^2-c_{\text{eff}}^2} \right), \label{eq:oblique_height_F} \\
	H_S &=& \frac{2}{W_S} \frac{Q}{\gamma_0} \left( \frac{c_S^2-v_A^2}{2c_{S}^2-v_A^2-c_{\text{eff}}^2} \right). \label{eq:oblique_height_S}
	\end{eqnarray}
	\label{eq:oblique_height}
\end{subequations}
Physically, the entropy mode does not contain a magnetic component; thus, the central Rayleigh peak remains unchanged in MHD. The emergence of additional peaks and their characteristics can also be explained physically. Specifically, these peaks correspond to two distinct MHD modes: the fast magnetosonic wave, propagating with a speed $c_F$, and the slow magnetosonic wave, propagating with a speed $c_S$. The frequency position of the fast magnetosonic mode is always greater than that for the slow magnetosonic mode. This is because of the effective equilibrium pressure for the fast magnetosonic wave being enhanced by additional magnetic pressure. The amount of enhancement depends on the oblique scattering angle.

\begin{figure*}
	\centering
	\begin{subfigure}[b]{0.485\textwidth}
		\centering
		\includegraphics[width=\textwidth,height=0.8\textwidth]{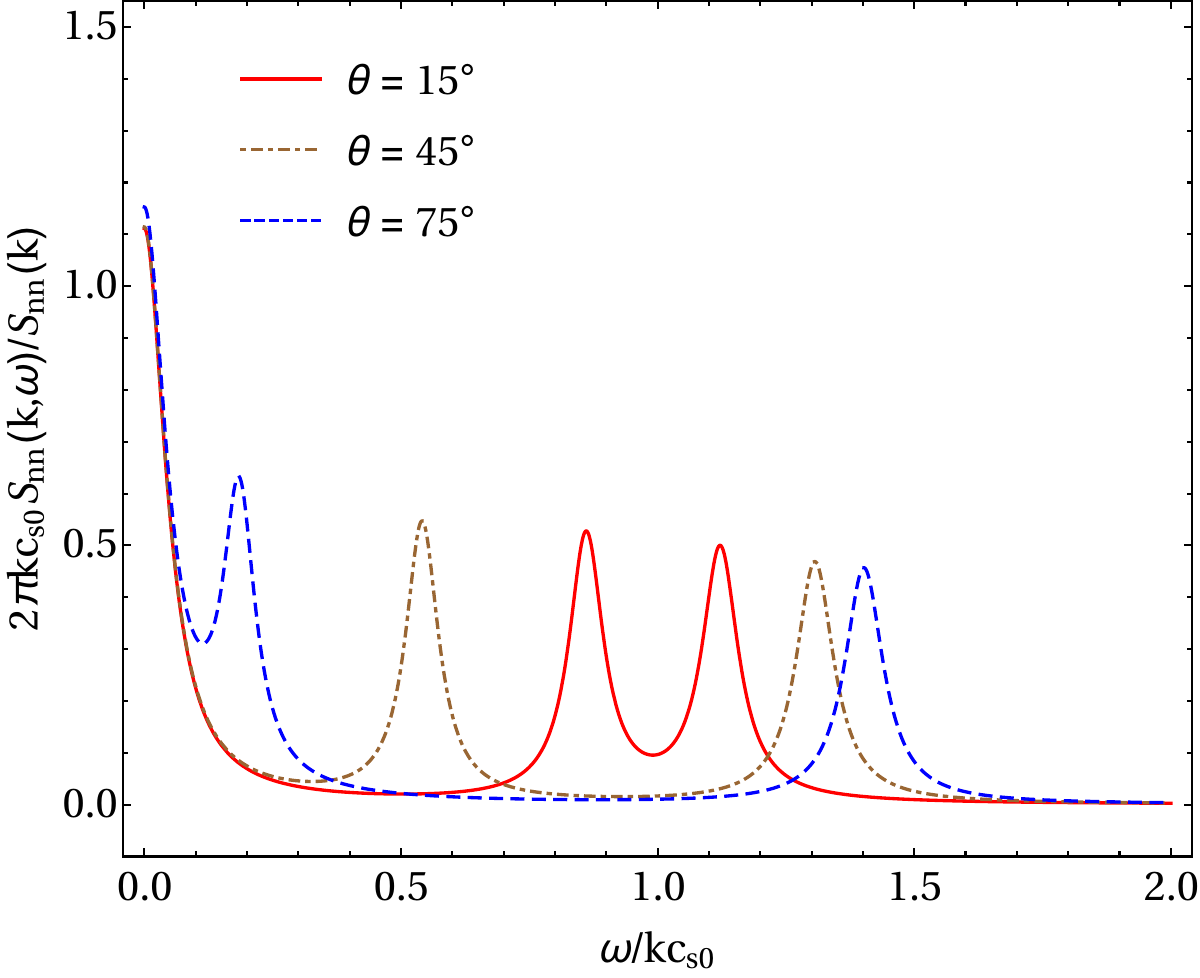}
		\caption{$c_Q/c_{s0}=0,$ and $v_A/c_{s0}=1.$}
		\label{fig:increasingObliquity_1}
	\end{subfigure}
	\hfill
	\begin{subfigure}[b]{0.485\textwidth}
		\centering
		\includegraphics[width=\textwidth,height=0.8\textwidth]{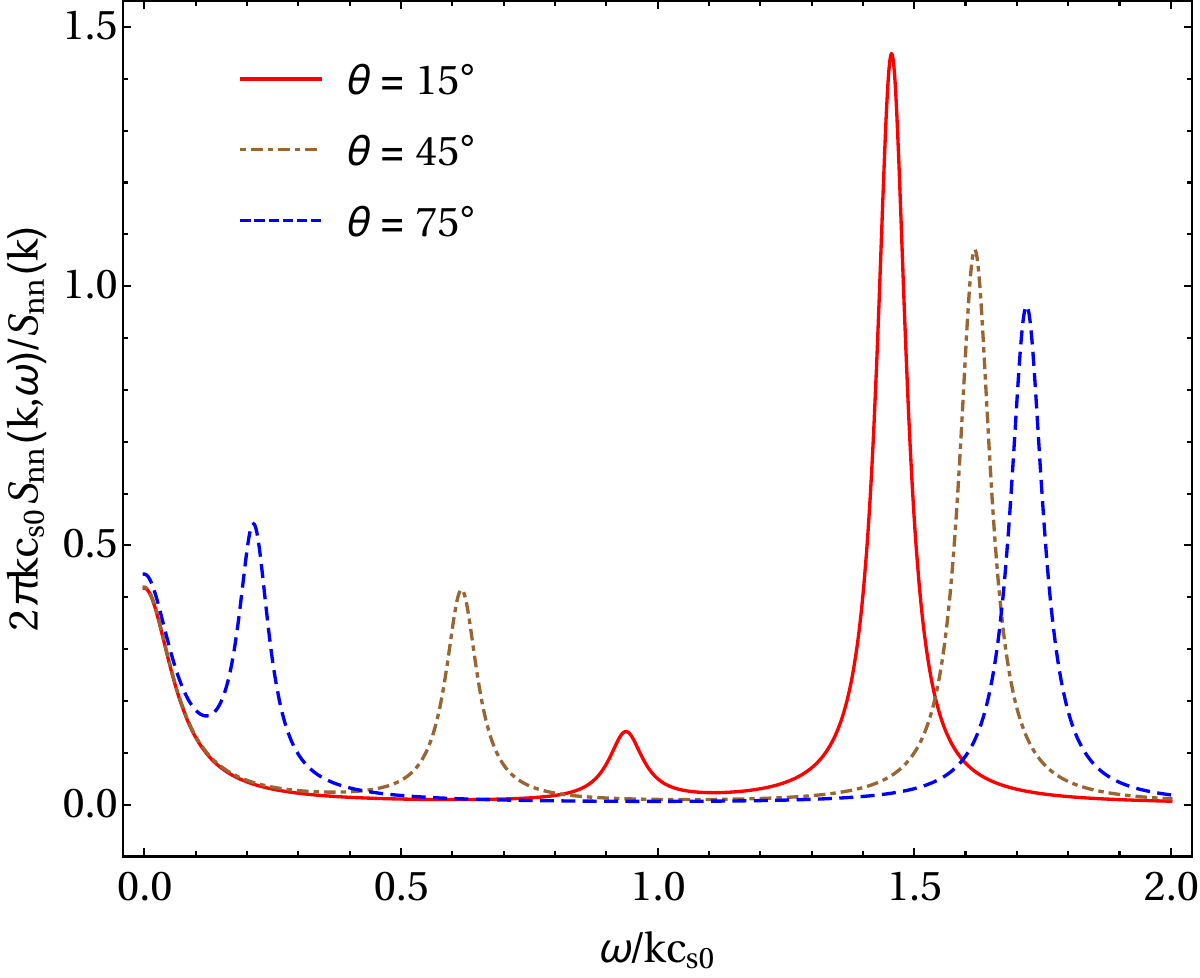}
		\caption{$c_Q/c_{s0}=1,$ and $v_A/c_{s0}=1.$}
		\label{fig:increasingObliquity_2}
	\end{subfigure}
	\caption{\textit{The dynamic structure factor in magnetized, high density plasma at different oblique scattering angles.} The plotted dynamic structure factors are calculated in the same way as in Fig. \ref{fig:parallel_1}, with the same dimensionless values for the dissipative terms.}
	\label{fig:increasingObliquity}
\end{figure*}
First, we consider a classical MHD system in which the equilibrium thermal and magnetic energy densities are comparable; this is equivalent to $v_A = c_{s0}$. In the classical MHD case, the speed associated with the quantum pressure vanishes (i.e., $c_Q = 0$). For quasiperpendicular perturbations (i.e., $\cos \theta \ll 1$), the thermal and magnetic pressure fluctuations are in phase for the fast magnetosonic waves. On the other hand, the slow magnetosonic waves become almost incompressible with thermal and magnetic pressure fluctuations acting out of phase. For the quasiparallel mode (i.e., $\cos \theta \lesssim 1$), the frequencies of the fast and slow magnetosonic modes are of similar orders of magnitude at a given wave number, but the fast mode's frequency is always greater. At positive frequencies, the positions of the fast and slow mode's peak become increasingly seperated as $\theta$ is increased. These points are illustrated in Fig.~\ref{fig:increasingObliquity_1}, for a fixed magnetization $(v_A = c_{s0})$ with increasing oblique scattering angles.

In the quantum MHD scenario, we start our analysis by considering the equilibrium quantum speed equating to the equilibrium sound speed, denoted as $c_Q = c_{s0}$. Furthermore, we assume $v_A = c_{s0}$. The resultant DSF is depicted in Fig.~\ref{fig:increasingObliquity_2}. In comparison to the classical MHD case, due to the introduction of Bohm contributions, the DSF exhibits modified peaks. Quantum effects enhance the effective thermal pressure for both magnetosonic waves. Consequently, the peak positions associated with fast and slow magnetosonic waves shift towards higher frequencies relative to their classical MHD counterparts. Another notable similarity is the increasing separation of peak positions linked to the fast and slow modes as the obliquity angle $\theta$ is increased. This behavior primarily stems from the fast mode's peak shifting to a higher frequency and the slow mode's peak shifting to a lower frequency as $\theta$ increases, in accordance with \Eq{eq:cF_cS}. 
\begin{figure}
	\centering
	\includegraphics[width=\columnwidth]{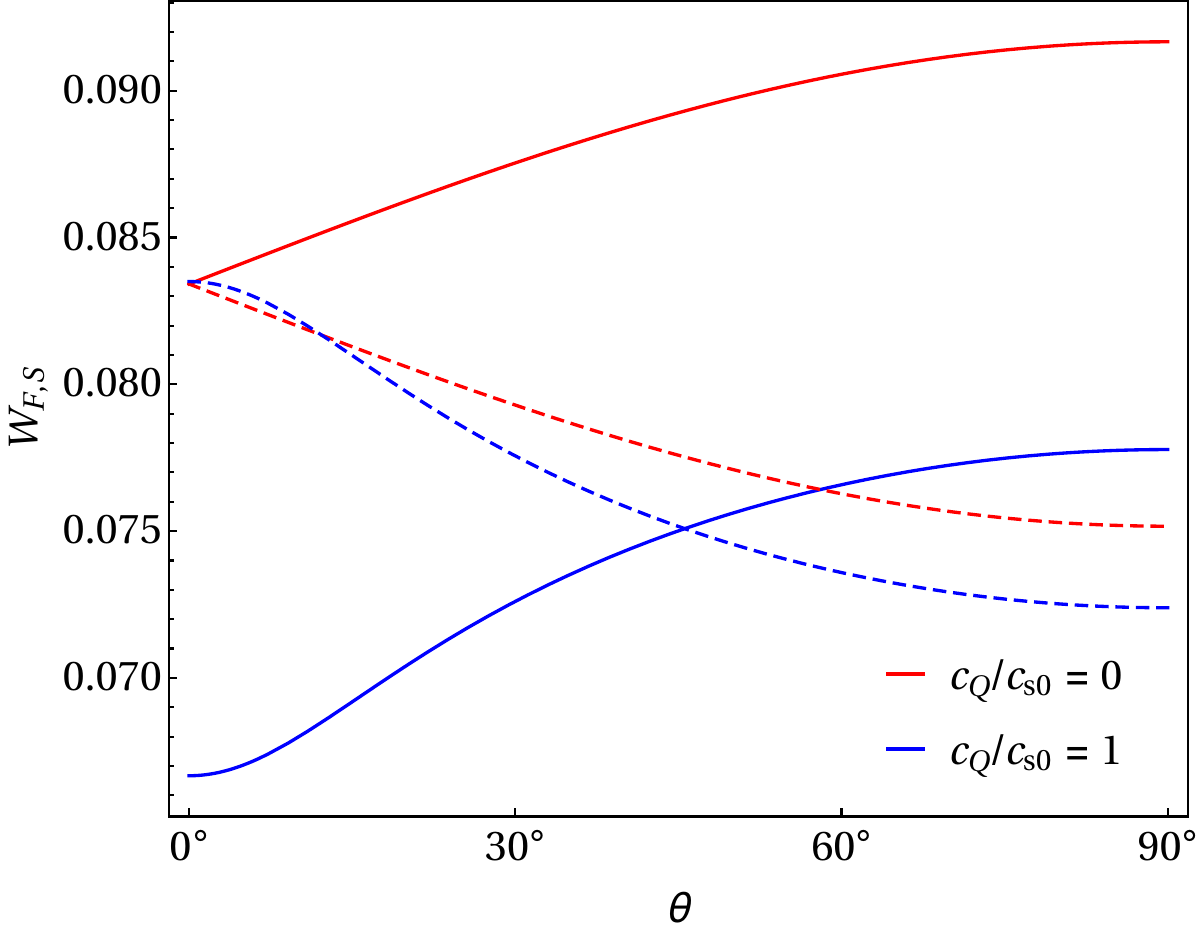}
	\caption{(Color online) \textit{The FWHM of fast (solid curves) and slow (dashed curves) magnetosonic modes} in magnetized, high density plasma with increasing oblique scattering angles for classical (red curves) and quantum (blue curves) MHD scenarios. We choose $v_A / c_{s0} = 1$. The magnitudes of various dissipative terms are consistent with those in Fig. \ref{fig:parallel_1}.}
	\label{fig:width_ObliqueAngle}
\end{figure}
The modification of $c_F$ and $c_S$ with $\theta$ directly impacts peak widths through their influence on the corresponding $\Gamma_{F,S}$ (Eq.~\ref{eq:oblique_diffuse}). For a given magnetic and quantum pressure, the coefficients associated with various dissipative terms in the expressions for $\Gamma_{F,S}$ change with the opposite sign as the oblique angle changes. Specifically, for the fast mode, the coefficients tied to thermal diffusivity and magnetic diffusivity diminish with increasing oblique angle, while the coefficient connected to compressive viscosity increases with rising $\theta$. Consequently, the width of the fast mode peak may either expand or contract, depending on the relative magnitudes of these dissipative terms. For a given set of dissipative terms, the width of the peak linked to the fast mode increases with the oblique angle, as shown in \Fig{fig:width_ObliqueAngle}. Conversely, the slow mode experiences a reduction in peak width due to the declining coefficients associated with thermal diffusivity, magnetic diffusivity, and compressive viscosity as $\theta$ increases. This is also illustrated in \Fig{fig:width_ObliqueAngle}.

In addition to changes in peak width, the peak heights corresponding to the fast and slow modes exhibit diverse variations as the oblique angle $\theta$ changes, while maintaining constant magnetic and quantum pressure conditions, as illustrated in \Fig{fig:increasingObliquity}. The peak height is primarily governed by two factors: (a) an inverse relationship with the peak width, and (b) a dependence on the velocity associated with the corresponding mode, as per \Eq{eq:oblique_height}. Consequently, as $\theta$ increases under fixed magnetic and quantum pressure, the peak height decreases for the fast mode and increases for the slow mode. This behavior remains consistent in both classical (Fig.~\ref{fig:increasingObliquity_1}) and quantum (Fig.~\ref{fig:increasingObliquity_2}) MHD cases. However, the quantum effects introduce further modifications through the `$Q$'-factor, which we will discuss in the following paragraph.

\begin{figure*}
	\centering
	\begin{subfigure}[b]{0.485\textwidth}
		\centering
		\includegraphics[width=\textwidth,height=0.8\textwidth]{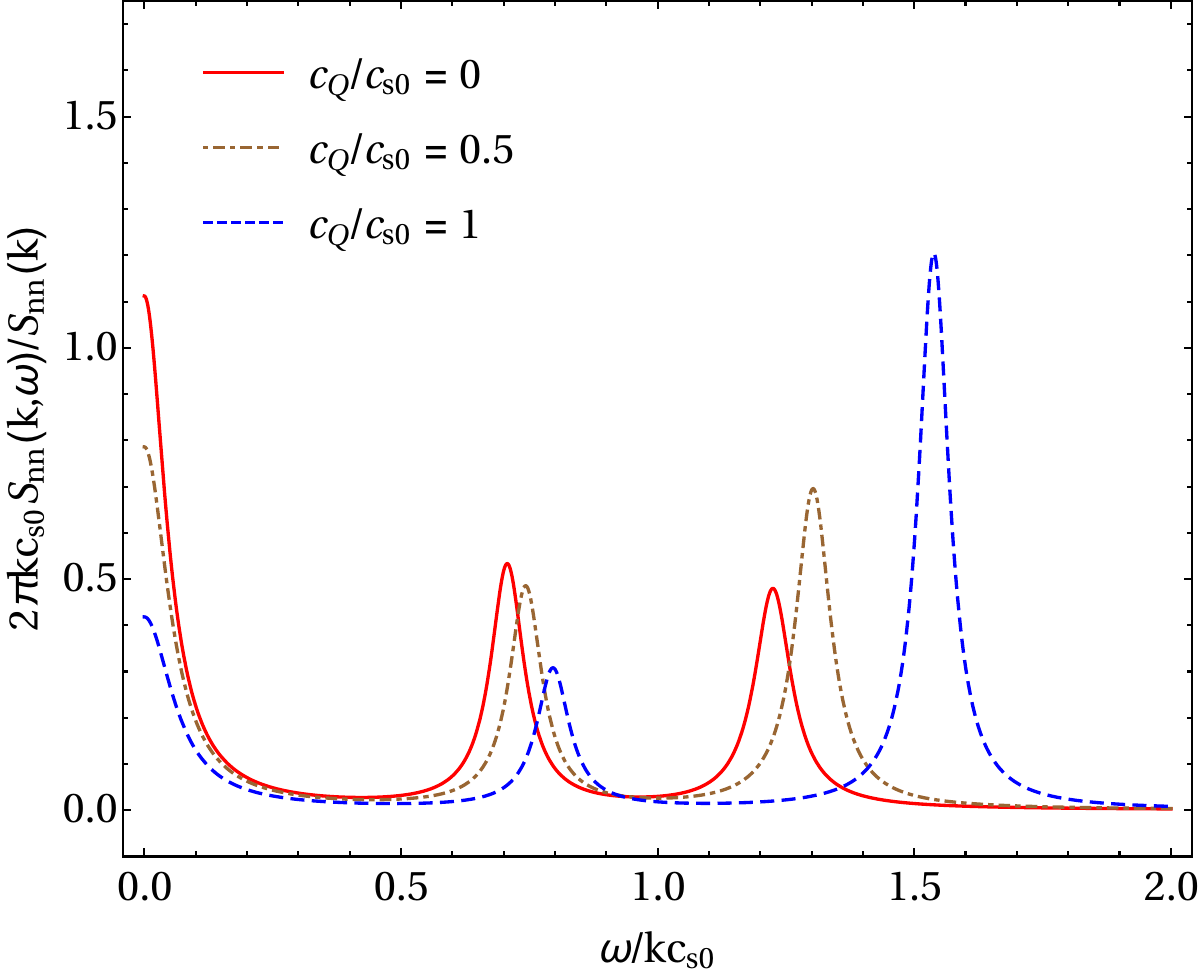}
		\caption{$v_A/c_{s0}=1,$ and $\theta=30^{\circ}$.}
		\label{fig:increasingQuantum_1}
	\end{subfigure}
	\hfill
	\begin{subfigure}[b]{0.485\textwidth}
		\centering
		\includegraphics[width=\textwidth,height=0.8\textwidth]{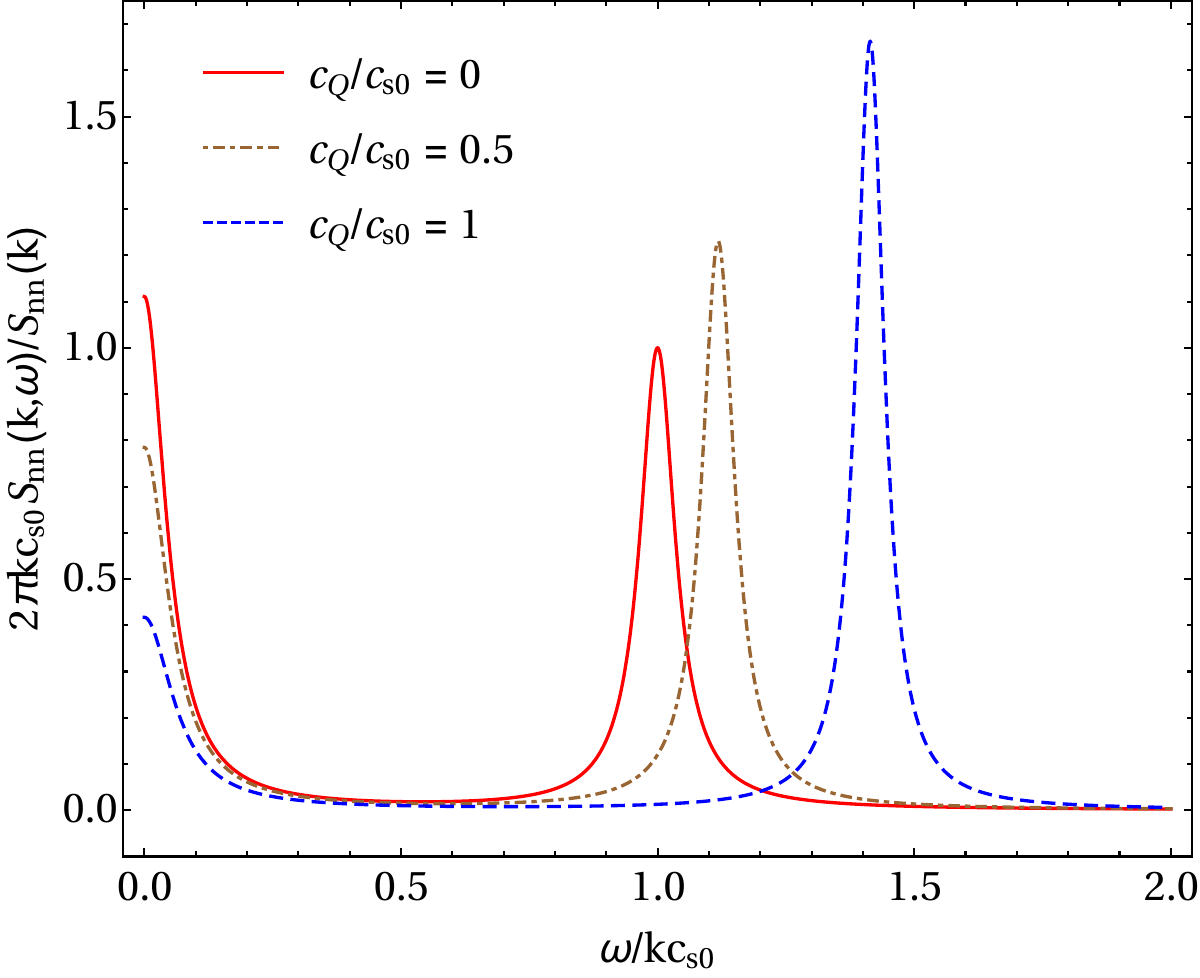}
		\caption{$v_A/c_{s0}=0,$ and $\theta \rightarrow$ no role.}
		\label{fig:increasingQuantum_2}
	\end{subfigure}
	\caption{\textit{The dynamic structure factor in magnetized, high density plasma with increasing quantum effects.} The plotted dynamic structure factors are calculated in the same way as in Fig. \ref{fig:parallel_1}, with the same dimensionless values for the dissipative terms.}
	\label{fig:increasingQuantum}
\end{figure*}
In \Fig{fig:increasingQuantum}, we provide an analysis of the influence of quantum effects on the DSF under the conditions of a constant magnetic pressure and an oblique angle. \Fig{fig:increasingQuantum_1} correcspons to the case $v_A/c_{s0}=1,$ and $\theta=30^{\circ}$, while \Fig{fig:increasingQuantum_2} is for $v_A/c_{s0}=0$. The non-magnetic scenario (Fig.~\ref{fig:increasingQuantum_2}) agrees with quantum hydrodynamics \cite{2012PhRvE..85d6408S}, with no dependence on the oblique angle $\theta$, thus yielding a DSF identical to that for parallel fluctuations, as illustrated in \Fig{fig:parallel}.

\begin{figure*}
	\centering
	\begin{subfigure}[b]{0.485\textwidth}
		\centering
		\includegraphics[width=\textwidth,height=0.8\textwidth]{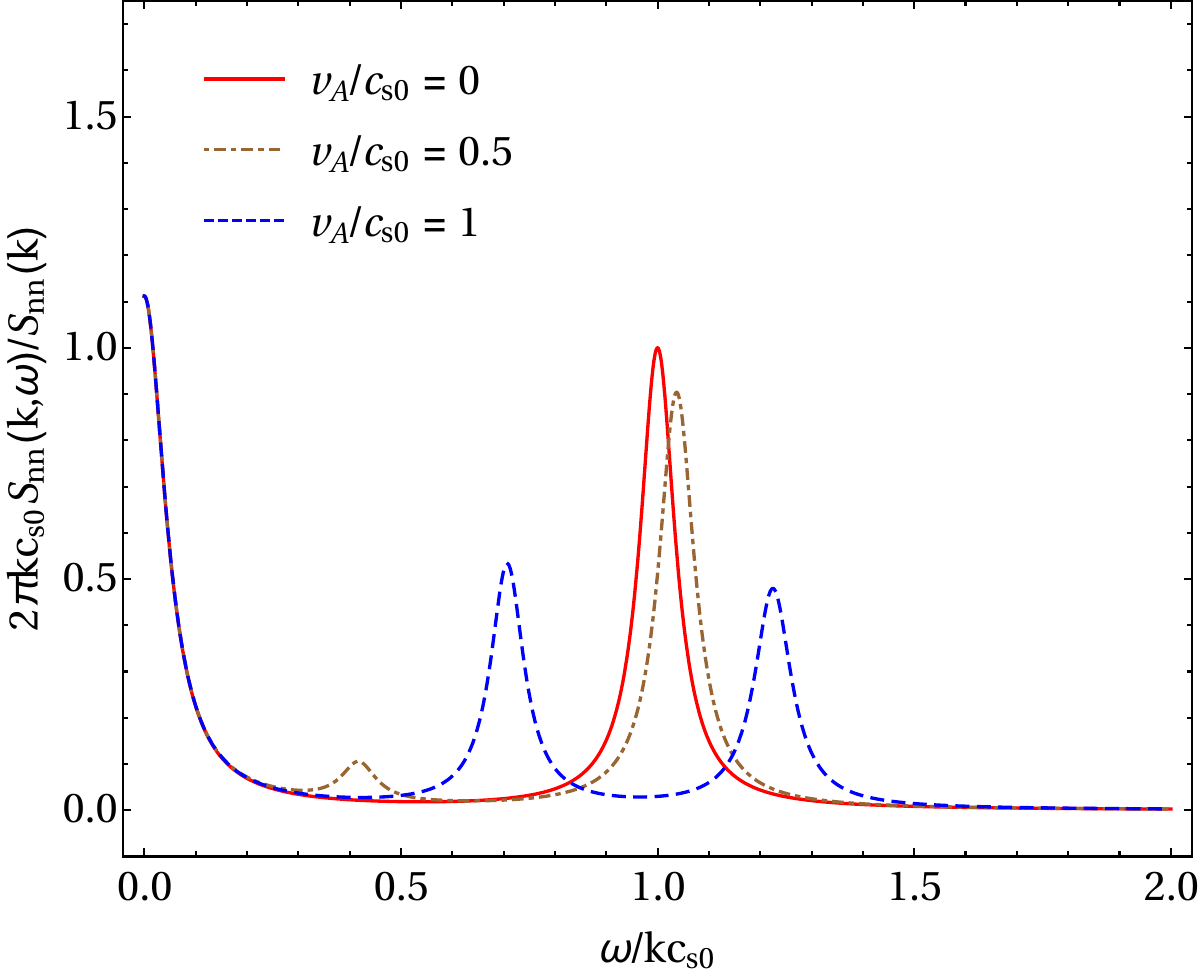}
		\caption{$c_Q/c_{s0}=0,$ and $\theta=30^{\circ}$.}
		\label{fig:increasingMagnetization_1}
	\end{subfigure}
	\hfill
	\begin{subfigure}[b]{0.485\textwidth}
		\centering
		\includegraphics[width=\textwidth,height=0.8\textwidth]{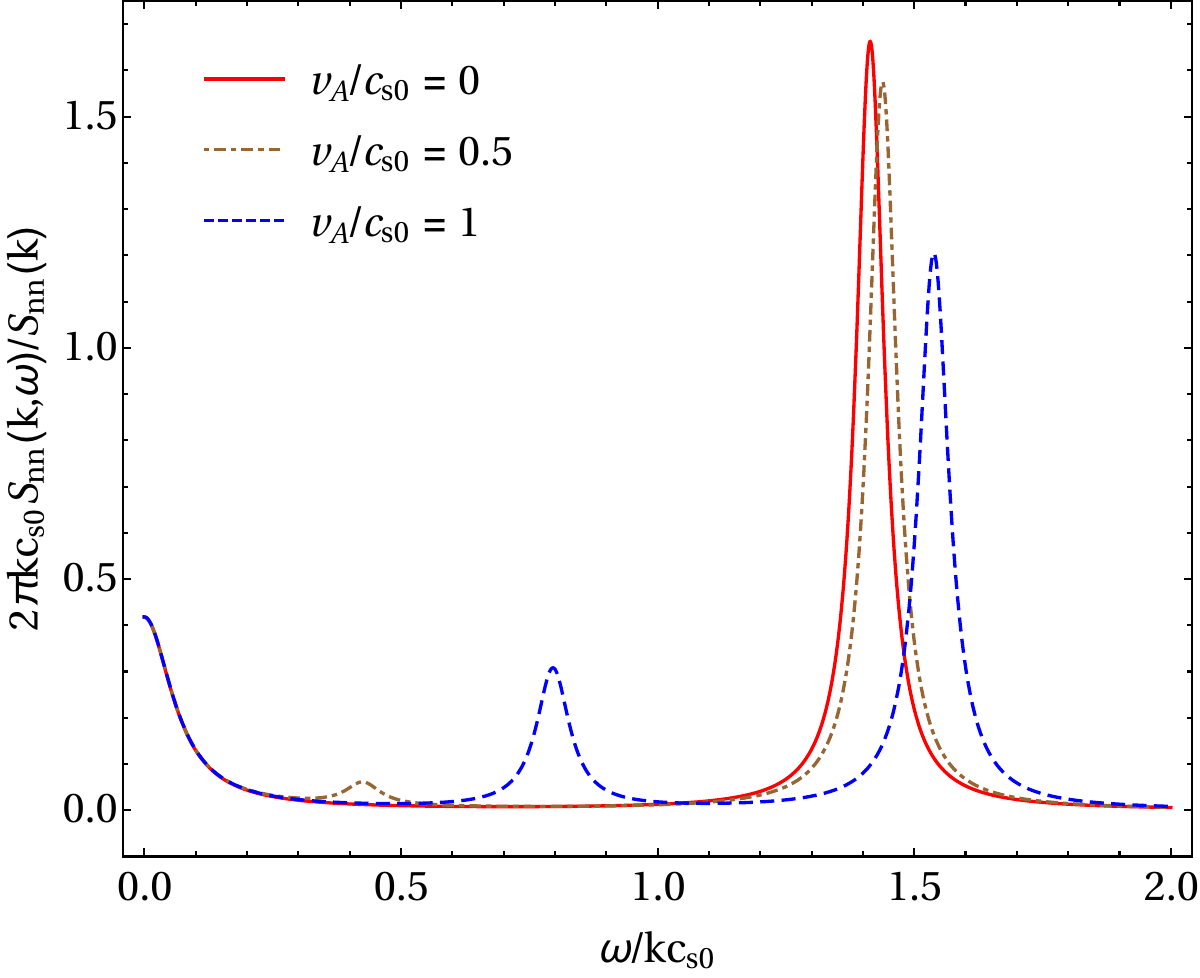}
		\caption{$c_Q/c_{s0}=1,$ and $\theta=30^{\circ}$.}
		\label{fig:increasingMagnetization_2}
	\end{subfigure}
	\caption{\textit{The dynamic structure factor in magnetized, high density plasma with increasing magnetization}. The plotted dynamic structure factors are calculated in the same way as in Fig. \ref{fig:parallel_1}, with the same dimensionless values for the dissipative terms.}
	\label{fig:increasingMagnetization}
\end{figure*}

Within the framework of quantum MHD, as depicted in \Fig{fig:increasingQuantum_1}, the DSF exhibits five peaks, as previously discussed. The frequency position of the central Rayleigh peak remains unaltered with variations in quantum pressure. Quantum effects, however, enhance the effective thermal pressure for both magnetosonic waves. Consequently, the frequency positions associated with the fast and slow modes shift towards higher frequencies with increasing quantum pressure, as described before. In addition, the Bohm contributions introduce two significant modifications: (a) \textit{a reduction of the adiabatic index by a factor of $Q$,} i.e., $\gamma'_0 = \gamma_0 / Q$, and (b) \textit{an enhancement of the effective thermal diffusivity by the same factor $Q$,} i.e., $\chi'_0 = Q\chi_0$. Consequently, the product $\gamma_0 \chi_0$ remains invariant. This modification is also observed in quantum hydrodynamics (i.e., with no magnetic field present). The increased thermal diffusivity due to quantum effects leads to the broadening of the FWHM of the Rayleigh peak through $\Gamma_\chi$, resulting in a reduction in peak height, as explained by \Eq{eq:oblique_height_R}. The width and height of both magnetosonic peaks are determined by a complex interplay of thermal diffusivity, resistivity, and viscosities, following \Eqs{eq:oblique_width}{eq:oblique_height}, respectively. Similarly, the DSF structures for classical and quantum MHD fluctuations are illustrated in \Figs{fig:increasingMagnetization_1}{fig:increasingMagnetization_2}, respectively, for a fixed oblique scattering angle $(\theta = 30^\circ)$ with increasing magnetization $v_A / c_{s0}$.

The total integrated intensity of the Rayleigh peak and each of the F- and S-mode peaks are, respectively,
\begin{subequations}
	\begin{eqnarray}
	I_R & = & \left[{\gamma_0-1 \over \gamma_0}{c_{s0}^2 \over c_{\text{eff}}^2}\right] S_{nn}(k) = \left[ 1-\frac{Q}{\gamma_0} \right] S_{nn}(k), \qquad \\
	I_F & = & {1 \over 2} \left[ \frac{Q}{\gamma_0} \left( \frac{c_F^2-v_A^2}{2c_{F}^2-v_A^2-c_{\text{eff}}^2} \right) \right] S_{nn}(k) , \qquad \\
	I_S & = & {1 \over 2} \left[ \frac{Q}{\gamma_0} \left( \frac{c_S^2-v_A^2}{2c_{S}^2-v_A^2-c_{\text{eff}}^2} \right) \right] S_{nn}(k)  . 
	\end{eqnarray}
\end{subequations}
Since, $(2c_{F}^2-v_A^2-c_{\text{eff}}^2) = -(2c_{S}^2-v_A^2-c_{\text{eff}}^2) = c_{F}^2 - c_{S}^2$, it preserves the sum rule, $I_R + 2I_F + 2I_S = S_{nn}(k)$. The ratio of the intensity of the central Rayleigh peak to that of the four shifted magnetosonic wave peaks is thus given by
\begin{equation}
\frac{I_R}{2I_F + 2I_S} = \left(\frac{\gamma_0}{Q} -1 \right) = (\gamma'_0 -1),
\end{equation}
and it follows the traditional Landau--Placzek ratio. Here, we introduce an additional ratio: the ratio of the intensity of the fast magnetosonic wave peak to that of the slow magnetosonic wave peak,
\begin{equation}
\frac{I_F}{I_S} = \frac{c_F^2-v_A^2}{v_A^2 - c_S^2} = \frac{c_F^2-v_A^2}{c_F^2 - c_{\text{eff}}^2}.
\label{eq:F-to-S}
\end{equation}
From an experimental point of view, we note that the \textit{F-to-S ratio} (Eq.~\ref{eq:F-to-S}) provides simultaneous sensitivity on both the magnetic field and the effective sound speed, which includes the Bohm contribution, in a magnetized, high-density plasma.

\section{Conclusions} \label{sec: conclusions}

In this paper, we have derived an analytical expression for the dynamic structure factor in a nonrelativistic, magnetized, high-density quantum plasma. Our approach involves describing collective excitations through the framework of magnetohydrodynamics, in which nonlocal quantum behavior is accounted for through the phenomenological Bohm potential. The inclusion of quantum effects is shown to have noticeable impacts in both hydrodynamics and magnetohydrodynamics conditions. Specifically, the Bohm contributions introduce three significant modifications: (a) \textit{an enhancement of the effective thermal pressure}, (b) \textit{a reduction of the adiabatic index by a factor of $Q$,} i.e., $\gamma'_0 = \gamma_0 / Q$, and (c) \textit{an enhancement of the effective thermal diffusivity by the same factor $Q$,} i.e., $\chi'_0 = Q\chi_0$. It is noteworthy that our analysis leads to the recovery of the same DSF structure as observed in standard classical hydrodynamic fluctuations when $Q=1$.

In the quantum hydrodynamic case, the DSF is shown to have the same three-peak structure---one central Rayleigh peak and two Brillouin peaks---as in the case of classical hydrodynamic fluctuations, but additional factors related to the quantum Bohm potential are now affecting their position, width, and intensity. The central Rayleigh peak is associated with non-propagating entropy fluctuations, thus its frequency position remains unchanged. The width of the Rayleigh peak is determined solely by thermal diffusivity through the parameter $\Gamma_\chi = Q\chi_0$, while its height is influenced by both $\gamma'_0$ and $\Gamma_\chi$ (Eq.~\ref{eq:height_parallel}). For the Brillouin peaks, quantum effects shift the frequency position to a higher frequency due to the enhanced effective thermal pressure from additional quantum pressure. The width of the Brillouin peaks is determined by both thermal diffusivity and viscosity, characterized by the parameter $\Gamma_{\|}$ (Eq.~\ref{eq:Gamma_parallel}), while their height is influenced by $\gamma'_0$ and $\Gamma_{\|}$ (Eq.~\ref{eq:height_parallel}).

In the context of quantum magnetohydrodynamics, the DSF maintains a five-peak structure, featuring a central Rayleigh peak and two pairs of peaks associated with fast and slow magnetosonic waves, as seen in classical magnetohydrodynamic fluctuations \cite{2019PhRvE..99f3204B}. However, the quantum Bohm potential introduces significant alterations, impacting their characteristics, including position, width, and intensity. In MHD, the structure of the DSF is contingent on the angle of fluctuations relative to the prevailing magnetic field within the medium. For fluctuations parallel to the magnetic field, the DSF retains its hydrodynamic nature, as expected. However, oblique fluctuations introduce an extra pair of peaks, associated with magnetic field fluctuations coupled with density fluctuations. Notably, both magnetosonic waves possess significant magnetic and thermal components. The enhancement of effective thermal pressure due to the quantum Bohm potential leads to a shift in the frequency positions of both pairs of peaks towards higher frequencies. Concurrently, resistive, viscous, and conductive dissipative processes dampen fast and slow magnetosonic waves. These damping factors determine the peak width through the respective parameter $\Gamma_{F,S}$, the general expression for which is provided in \Eq{eq:oblique_diffuse}. The height of these peaks is contingent upon their width (inversely proportional), the modified adiabatic index shaped by quantum effects, and the relative propagation speeds of various waves, following \Eq{eq:oblique_height}. Finally, the central Rayleigh peak, representing the zero-frequency, non-propagating entropy mode and devoid of any magnetic component, remains unaltered in MHD.

\begin{table*}
	\centering
	\caption{Parameters and quantities for plasma diagnostics using the DSF of a magnetized, high-density plasma.}
	\label{tab:plasma-diagnostics}
	\begin{tabular}{|p{5.5cm}|p{5cm}|p{5cm}|}
		\hline
		\textbf{Parameter/Quantity from DSFs} & \textbf{Description} & \textbf{Plasma Properties} \\
		&	&	\\
		\hline
		Peak positions & Associated with two magnetosonic waves	& Values for $c_F$ and $c_S$ \\
		\hline
		Landau-Placzek ratio ($R_{LP}$) & $ I_R / (2I_F + 2I_S) = (\gamma'_0 - 1)$ & Adiabatic index $(\gamma'_0)$ \\
		&	&	\\
		\hline
		F-to-S ratio ($R_{FS}$) & $I_F / I_S = (c_F^2-v_A^2)/(v_A^2 - c_S^2)$ & Alfv\`en speed ($v_A$), and hence, \\ 	& 	& magnetic field strength \\
		\hline
		Identity I (obtained from Eq.~\ref{eq:cF_cS}) & $c_F^2 + c_S^2 = v_A^2 + c_{\text{eff}}^2$ & Speed associated with the effective thermal pressure ($c_{\text{eff}}$) \\
		\hline
		Identity II & $c_F^2 c_S^2 = v_A^2 c_{\text{eff}}^2 \cos^2{\theta}$ &  Oblique angle ($\theta$) \\
		(obtained from Eq.~\ref{eq:cF_cS}) 	&		& 		\\
		\hline
		Width of the central Rayleigh peak  & $\Gamma_{\chi} = \chi'_0 $ & Thermal diffusivity ($\chi'_0$) \\
		via $\Gamma_\chi$ 	&	& \\
		\hline
		Width of the fast and slow mode peaks via $\Gamma_{F,S}$ & Expressions given in Eqs.~(\ref{eq:oblique_diffuse_1}) and (\ref{eq:oblique_diffuse_2}) & Correlations$^*$ between resistivity ($\eta_0$) and viscosities ($\nu_{s0}$ and $\nu_{c0}$) \\
		\hline
	\end{tabular}
	\medskip 
	\caption*{\small $^*$ \textbf{Note:} It is important to note that the three unknown transport coefficients---resistivity, bulk viscosity and shear viscosity---cannot be simultaneously determined from two known equations (Eqs.~\ref{eq:oblique_diffuse_1} and \ref{eq:oblique_diffuse_2}), unless one of these coefficients is known to be small.}
\end{table*}
From an experimental perspective, the presence of multiple peaks in the DSF offers a unique opportunity for the simultaneous measurement of various plasma properties, as well as transport and thermodynamic coefficients in magnetized laboratory plasmas. An important parameter connecting theoretical predictions and experimental observations is the Landau-Placzek ratio, denoted as $R_{LP} = I_R / 2I_B = (\gamma_0 - 1)$. This ratio was originally derived from a Rayleigh-Brillouin triplet and provides a direct means to determine the specific heat ratio of diverse liquids and gases through experiments \cite{1966JChPh..44.2785C, 1967JChPh..47...31O, 2011JChPh.135m4510P, 2017PhRvE..96d2608Z}. In the context of quantum hydrodynamics, we have found that this ratio modifies to $R_{LP} = I_R / 2I_B = (\gamma'_0 - 1)$. For DSFs featuring a five-peak structure, as observed in magnetohydrodynamics, the equivalent form of the Landau-Placzek ratio becomes $ I_R / (2I_F + 2I_S) = (\gamma'_0 - 1)$. Additionally, we introduce another significant ratio known as the `\textit{F-to-S ratio},' denoted as $I_F / I_S = (c_F^2-v_A^2)/(v_A^2 - c_S^2)$. A suggested strategy for measuring various plasma properties is outlined sequentially in Table~\ref{tab:plasma-diagnostics}. Hence, the DSF can serve as a robust diagnostic tool for plasma properties that are otherwise very challenging to measure.

\begin{acknowledgments}

The work of GG was partially supported by the Science and Technology Facilities Council (STFC) grant no. ST/W000903/1. 

\end{acknowledgments}

\appendix
\section{Thermodynamic Identities} \label{sec: Appendix-A}

In this Appendix, we derive the temperature evolution equation (Eq.~\ref{MHDgoveqns_temp}) from the internal energy equation (Eq.~\ref{eq:energy}). Here, we also derive \Eq{thermoiden_B} for pressure in terms of state variables density and temperature. Using the first law of thermodynamics and the continuity equation (Eq.~\ref{eq:continuity}), we write down the internal energy per unit mass as
\begin{equation}
\frac{d\epsilon}{dt} = T \frac{dS}{dt} + \frac{p}{\rho^2} \frac{d\rho}{dt} = T \frac{dS}{dt} - \frac{p}{\rho} \div \bm{u},
\label{eq:E_to_S}
\end{equation}
where $S$ represents the specific entropy. Substituting \Eq{eq:E_to_S} into \Eq{eq:energy}, we obtain a conservation law for specific entropy, as given by 
\begin{equation}
\rho T \frac{dS}{dt} = -\bm{\Pi}:\bm{\nabla}\bm{u} + \eta\frac{\vert\curl{\bm{B}}\vert^{2}}{\mu_{0}} - \div{\bm{q}} + \bm{\Phi}_{\text{Bohm}}\cdot\bm{u}. \label{eq:entropy}
\end{equation}
Assuming that the specific entropy $S = S(\rho, T)$, the total differential is given by
\begin{equation}
dS = \left(\frac{\partial S}{\partial \rho}\right)_T d\rho + \left(\frac{\partial S}{\partial T}\right)_\rho dT.
\end{equation}
Applying Maxwell’s identities and the first law of thermodynamics, we determine
\begin{equation}
\left(\frac{\partial S}{\partial \rho}\right)_T = - \frac{1}{\rho^2}\left(\frac{\partial p}{\partial T}\right)_\rho = \frac{C_V - C_P}{\alpha_T \rho T}, \ \left(\frac{\partial S}{\partial T}\right)_\rho = \frac{C_V}{T},
\end{equation}
where $C_V$ is the heat capacity at constant volume, $C_P$ is the heat capacity at constant pressure, and $\alpha_T \equiv -\rho^{-1} (\partial \rho / \partial T)_p$ is the coefficient of thermal expansion. We obtain 
\begin{equation}
dS = \frac{C_V}{T} \left( dT - \frac{\gamma - 1}{\alpha_T \rho} d\rho \right),
\end{equation}
where we have used $\gamma = C_P / C_V$. Using the continuity equation (Eq.~\ref{eq:continuity}), we can finally write
\begin{equation}
\frac{dS}{dt} = \frac{C_V}{T} \left( \frac{dT}{dt} - \frac{\gamma - 1}{\alpha_T \rho} \frac{d\rho}{dt} \right) = \frac{C_V}{T} \left( \frac{dT}{dt} + \frac{\gamma - 1}{\alpha_T } \div \bm{u} \right).
\label{eq:entropy2}
\end{equation}
Substituting \Eq{eq:entropy2} into \Eq{eq:entropy} yields the desired temperature evolution equation (Eq.~\ref{MHDgoveqns_temp}).

Similarly, for the pressure $p = p(\rho, T)$, we determine the total differential
\begin{equation}
dp = \left(\frac{\partial p}{\partial \rho}\right)_T d\rho + \left(\frac{\partial p}{\partial T}\right)_\rho dT.
\end{equation}
Using reciprocity and Maxwell’s identities, we obtain
\begin{eqnarray}
\left(\frac{\partial p}{\partial \rho}\right)_T &=& \frac{C_V}{C_P} \left(\frac{\partial p}{\partial \rho}\right)_S = \frac{c_s^2}{\gamma}, \nonumber \\
\left(\frac{\partial p}{\partial T}\right)_\rho &=& - \left(\frac{\partial p}{\partial \rho}\right)_T \left(\frac{\partial \rho}{\partial T}\right)_p = \frac{c_s^2}{\gamma} \rho \alpha_T,
\end{eqnarray}
where $c_s^2 \equiv \left(\partial p / \partial \rho \right)_S$ represents the adiabatic sound speed. These expressions lead to the final expression
\begin{equation}
dp = \frac{c_s^2}{\gamma} (d\rho + \rho \alpha_T \ dT),
\end{equation}
from which \Eq{thermoiden_B} follows trivially.



%


\end{document}